\documentclass{elsart}

\journal{Astroparticle Physics}
\usepackage{amsmath,amssymb}
\usepackage{graphicx}
\usepackage{epsfig}
\usepackage{color}
\usepackage{verbatim}

\def\gs{\mathrel{
   \rlap{\raise 0.511ex \hbox{$>$}}{\lower 0.511ex \hbox{$\sim$}}}}
\def\ls{\mathrel{
   \rlap{\raise 0.511ex \hbox{$<$}}{\lower 0.511ex \hbox{$\sim$}}}}

\newcommand{\ba}{\begin{array}{c}}
\newcommand{\baz}{\begin{array}{cc}}
\newcommand{\bad}{\begin{array}{ccc}}
\newcommand{\bav}{\begin{array}{cccc}}
\newcommand{\baf}{\begin{array}{ccccc}}
\newcommand{\bena}{\begin{eqnarray}}
\newcommand{\eena}{\end{eqnarray}}
\newcommand{\bea}{\begin{equation} \begin{array}{c}}
\newcommand{\eea}{ \end{array} \end{equation}}
\newcommand{\ea}{\end{array}}

\begin{document}

\begin{flushright}
{YITP-SB-13-34 \\
UB-ECM-PF-95-13 \\
ICCUB-13-232
}
\end{flushright}

\begin{frontmatter}

\title{Reevaluation of the Prospect of Observing Neutrinos from Galactic Sources in the Light 
of Recent Results in Gamma Ray and Neutrino Astronomy}

\author[a,b,c]{M.~C.~Gonzalez-Garcia}
\ead{concha@insti.physics.sunysb.edu}
\author[d]{F.~Halzen}
\ead{halzen@icecube.wisc.edu}
\author[b]{V.~Niro}
\ead{niro@ecm.ub.edu}

\address[a]{Instituci\'o Catalana de Recerca i Etudes Avan\c{c}ats (ICREA)}
\address[b]{
  Departament d'Estructura i Constituents de la Mat\`eria and Institut de Ciencies del Cosmos, \\Universitat de
  Barcelona, Diagonal 647, E-08028 Barcelona, Spain}

\address[c]{C.N.~Yang Institute for Theoretical Physics, and 
Department of Physics and Astronomy, Stony Brook University,
\\ Stony Brook, NY 11794-3840, USA}

\address[d]{Wisconsin IceCube Particle Astrophysics Center and Department of Physics, \\
University of Wisconsin, Madison, WI 53706, USA}

\begin{abstract}
In light of the recent IceCube evidence for a flux of extraterrestrial
neutrinos, we revisit the prospect of observing the sources of the
Galactic cosmic rays. In particular, we update the predictions for the
neutrino flux expected from sources in the nearby star-forming region
in Cygnus taking into account recent TeV gamma ray measurements of
their spectra.  We consider the 
three Milagro sources: MGRO J2019+37,
MGRO J1908+06 and MGRO J2031+41 
 and calculate the attainable 
confidence level limits and statistical significance as a function 
of the exposure time.  
We also evaluate the prospects for a kilometer-scale detector
in the Mediterranean to observe and elucidate the origin of the cosmic
neutrino flux measured by IceCube.
\end{abstract}


\end{frontmatter}

\section{\label{sec:intro} Introduction}

If supernova remnants are indeed the sources of the highest energy
Galactic cosmic rays \cite{BaadeAndZwicky}, the IceCube neutrino
detector is expected to detect a flux of neutrinos accompanying the
observed cosmic ray flux. Fermi has recently established the presence
of pions in two supernova remnants thus unambiguously indicating the
acceleration of cosmic rays~\cite{Ackermann:2013wqa}. However, their
energies do not reach the PeV range and therefore the ``PeVatrons"
that are the sources of the cosmic rays in the ``knee" region of the
spectrum, and above, remain unidentified. Generic PeVatrons produce
pionic gamma rays whose spectrum extends to several hundred TeV
without a cutoff. Their predicted flux should be within reach of the
present generation of ground-based gamma ray telescopes but has not
been identified so far.

The highest energy survey of the Galactic plane to date has been
performed by the Milagro detector. In particular, the survey in the
10\,TeV band has revealed a subset of sources located 
within nearby
star-forming regions in Cygnus and in the vicinity of Galactic
latitude $l=40$\,degrees. Subsequently, directional air Cherenkov
telescopes were pointed at some of the
sources~\cite{DjannatiAtai:2007ny,Albert:2008yk}, revealing them as
PeVatron candidates with gamma-ray fluxes following an $E^{-2}$ energy
spectrum that extends to tens of TeV without evidence for a
cutoff. Interestingly, some of the sources cannot be readily
associated with known supernova remnants, or with any non-thermal
sources observed at other wavelengths. These are likely to be
molecular clouds illuminated by the cosmic-ray beam accelerated in
young remnants located within about 100~pc. Indeed one expects that
multi-PeV cosmic rays are accelerated only over a short time period
when the shock velocity is high, i.e., between free expansion and the
beginning of its dissipation in the interstellar medium. The
high-energy particles can produce photons and neutrinos over much
longer periods when they diffuse through the interstellar medium to
interact with nearby molecular clouds~\cite{Gabici:2007qb}. An
association of molecular clouds and supernova remnants is expected in
star-forming regions. Note that any confusion between pionic with
synchrotron photons is unlikely to be a problem in this case.

Assuming that the Milagro sources are indeed cosmic-ray accelerators,
the equality of the production of pions of all three charges dictates
the relation between pionic gamma rays and neutrinos and basically
predicts the production of a $\nu_\mu+\bar\nu_\mu$ pair for every two
gamma rays seen by Milagro. The calculation can be performed in more
detail with approximately the same
outcome~\cite{Halzen:2008zj,GonzalezGarcia:2009jc}.  For average
values of the source parameters it was anticipated that the completed
IceCube detector should confirm sources in the Milagro sky map as
sites of cosmic-ray acceleration at the $3\sigma$ level in less than
one year and at the $5\sigma$ level in three years. This assumes that
the source extends to 300\,TeV, i.e. approximately 10\% of the energy
of the cosmic rays near the knee in the spectrum. There are intrinsic
ambiguities of an astrophysical nature in this estimate that may
reduce or extend the time required for a $5\sigma$
observation~\cite{GonzalezGarcia:2009jc}, most prominently the exact
location where the sources run out of energy.  Also, the extended
nature of some of the Milagro sources represents a challenge for
IceCube observations that are optimized for point sources. 
For other previous analyses of galactic sources of high-energy neutrinos at IceCube, 
we refer also to Refs.~\cite{Vissani:2011ea} and~\cite{Vissani:2011vg}.

IceCube searches have revealed positive fluctuations from these
sources in the 8 years of AMANDA data and in 4 out of 5 years of data
collected with the partially deployed IceCube detector. On the other
hand, 
the first extraterrestrial neutrino flux observed~\cite{Aartsen:2013jdh}
by IceCube consists of 28 events (more below) with 
no event originating from the nearby star-forming region in Cygnus. 
This fact, together with the availability of new information from gamma ray 
telescopes, has motivated  us to revisit the calculation of the neutrino flux
in Ref.~\cite{GonzalezGarcia:2009jc}  for some of the sources. 
In particular we will update the information on MGRO J2019+37, MGRO
J1908+06 and MGRO J2031+41~\cite{ARGO-YBJ:2012goa,Abdo:2012jg}, 
3 of
the 6 sources used in the IceCube stacking analysis based on
references~\cite{GonzalezGarcia:2009jc,Halzen:2008zj,Kappes:2009zza}. 

Note that recently a lot of work has been done to try to explain 
the IceCube results in terms of point sources. 
For example, in Ref.~\cite{Fox:2013oza} the authors discuss the possibility to explain 
the IceCube data, and in particular an hot spot of 7 shower events, with 24 TeV unidentified sources 
of our Galaxy.
Among the sources considered, there are also the two Milagro sources, MGRO~J1908+06 
and MGRO~J2031+41, assumed to be Galactic hypernova remnant. Note that  MGRO J2019+37 is, instead, identified as pulsar wind 
nebulae (PWN). 
The conclusion of the analysis is that only 3.8 of the IceCube events may originate 
from the TeV unidenfied sources, conclusion obtained by comparing the spatial distribution of the 
IceCube events and the one from the unidentified sources. 
In Ref.~\cite{Kistler:2006hp}, instead, numerous Galactic sources are examined 
and for them the shower event rates and muon event rates are calculated. 
For example, Vela Jr. (RX J0852.0–4622) is a southern-sky source, 
that could be observed as muon tracks by 
an km$^3$ Northen hemisphere detector and through cascade events by the IceCube detector. 
Using both the muon tracks and cascades, it could be possible to better identify the specific source, pinning down 
its characteristics. In particular both the location (through muon events) and the source spectrum 
(through cascade events) could be reconstructed with precision. 

We want to stress, however, that the three Milagro sources that we are going to considere in this paper 
will give as main channel in IceCube muon tracks, since they are located in the Northern hemisphere. 
For this reason, this is the main event signal that we are going to calculate.
We won't consider shower events in a Northern hemiphere detector for these three Milagro 
sources. Our main scope is to consider the muon tracks and analyzing the possibility that 
these sources could be detected or not in less than 10 years at IceCube. However, we will 
instead consider a Northern hemisphere detector under the hypothesis of testing 
an hot spot recently revealed by IceCube, as described in the following. 

As mentioned above, recently, IceCube has presented the first evidence for an
extraterrestrial flux of very high-energy neutrinos, some with PeV
energies. IceCube has thus become the latest entry in an extensive and
diverse collection of instruments attempting to pinpoint the still
enigmatic sources of cosmic rays. Analyzing data 
collected between May 2010 and May 2012, 28 neutrino events were
identified with in-detector deposited energies between 30 and 1200
TeV.   Among the 28 events, 21
are showers whose energies are measured to better than 15\% but whose
directions are determined to $10-15$ degrees only. None show evidence
for a muon track accompanying the neutrino. If of atmospheric origin,
the neutrinos should be accompanied by muons produced in the air
shower in which they originate. For example, the probability that a
PeV atmospheric neutrino interacting in IceCube is unaccompanied by a
muon is of order 0.1\%. The remaining seven events are muon
tracks. With the present statistics, these are difficult to separate
from the competing atmospheric background.  Fitting the data to a
superposition of an extraterrestrial neutrino flux on an atmospheric
background yields a cosmic neutrino flux of
\begin{equation}
E_\nu^2 \frac{dN_\nu}{dE_\nu}=3.6\times10^{-11}\,\rm TeV\,cm^{-2}\,s^{-1}\,sr^{-1}
\label{eq:heseflux}
\end{equation}

Of those 28 events, a hot spot of 7 shower events is evident at 
RA=281 degrees and dec=23
degrees close to the Galactic center although its significance is only
8\% according to the test statistic defined in the blind analysis of
the IceCube data. On the other hand, the highest energy event does
reconstruct to within 1 degree of the Galactic center and, assuming an
isotropic distribution, only 0.7 events are expected in the area of
the sky covered by the seven showers. We will speculate that PeVatrons
producing cosmic rays in the $10^{15}-10^{17}$ energy range are the
origin of these neutrinos. It is not unreasonable to expect that
PeVatrons cluster in the direction of the Galactic center
corresponding to the largest concentration of mass along the line of
sight. The star-forming region near the Galactic center itself is
likely to be distant to be observed individually. We will investigate
the opportunities for a kilometer-scale detector in the Northern
hemisphere to elucidate the origin of the IceCube flux by observing
muon neutrinos which allow for sub-degree angular reconstruction.

In Sec.~\ref{sec:point_sources} we update the gamma ray spectra from
the three Milagro sources: MGRO J2019+37, MGRO J1908+06 and MGRO
J2031+41 and the calculation of the associated neutrino event rates in
IceCube. In Sec.~\ref{sec:results} we study the attainable confidence
level limits and statistical significance which Icecube can set as a
function of the exposure time considering different values of the
source parameters.  We also estimate the prospects for a
kilometer-scale detector in the Mediterranean to observe and elucidate
the origin of the cosmic neutrino flux measured by IceCube.  We
briefly summarize our conclusions in Sec.~\ref{sec:conclu}.

\section{\label{sec:point_sources} Point sources}

\subsection{\label{sec:neutrino_flux} Neutrino flux}
We start by updating the information on 3 of the 6 MILAGRO sources
considered in past IceCube analyses. The parameters in
Table~\ref{tab:sources}  refer to the parametrization reported in
Refs.~\cite{ARGO-YBJ:2012goa,Abdo:2012jg}, where the $\gamma$-ray flux
in the TeV energy range is parametrized in terms of a spectral slope
$\alpha_{\gamma}$, an energy $E_{cut,\gamma}$ where the accelerator
cuts off, and a normalization $k_{\gamma}$ as
\begin{equation}
\frac{dN_{\gamma}(E_\gamma)}{dE_\gamma}
=k_{\gamma}
 \left(\frac{E_\gamma}{\rm TeV}\right)^{-\alpha_{\gamma}} 
\exp\left(-\frac{E_\gamma}{E_{cut,\gamma}}\right)\,,
\label{eq:Ecutpaper}
\end{equation}
for a power law with cut-off fit and with
\begin{equation}
k_{\gamma}\equiv~\frac{{\rm K}}{(E_{\rm norm}/{\rm TeV})^{-\alpha_{\gamma}}}\,,
\label{eq:kgamma}
\end{equation}
where $E_{\rm norm}$ is the energy in TeV at which the flux is
normalized, {\it i.e.} for $E\equiv E_{\rm norm}$, the value of the
flux $dN_{\gamma}/dE_\gamma$ would be equivalent to K, in the absence
of an energy cut-off. For sources without a cut-off the parametrization used 
is the one in Eq.~(\ref{eq:Ecutpaper}) simply setting
$E_{cut,\gamma}\rightarrow \infty$. 

The neutrino fluxes associated with pionic gamma rays emitted by a source
is directly determined by particle physics; approximately
one $\nu_\mu + \bar\nu_\mu$ pair should accompany every 2 gamma rays.
The exact relation between the gamma ray and neutrino fluxes has been
described in detail in Ref.\cite{Kelner:2006tc,Kappes:2006fg}. Their
starting point, however, is a parametrization of
the gamma-ray flux slightly different from the one 
in Eq.(\ref{eq:Ecutpaper}), namely  
\begin{equation}
\frac{dN_{\gamma}(E_\gamma)}{dE_\gamma}
=k_{\gamma}
 \left(\frac{E_\gamma}{\rm TeV}\right)^{-\alpha_{\gamma}} 
\exp\left(-\sqrt{\frac{E_\gamma}{E_{cut,\gamma}}}\right)\,.
\label{eq:Ngamma}
\end{equation}  

Next, following   the approximate relations  given in  
Ref.\cite{Kelner:2006tc,Kappes:2006fg} we can write the corresponding
neutrino flux at the Earth after oscillations as 
\begin{equation} 
\frac{dN_{\nu_\mu+\bar\nu_\mu}(E_\nu)}{dE_\nu}
=k_{\nu} \left(\frac{E_\nu}{\rm TeV}\right)^{-\alpha_ {\nu}}
\exp\left(-\sqrt{\frac{E_\nu}{E_{cut,\nu}}}\right)\,,
\label{eq:flux_nu}
\end{equation} with
\begin{eqnarray}
&& k_\nu=(0.694-0.16 \alpha_\gamma) k_\gamma \nonumber\,, \\ &&
  \alpha_\nu= \alpha_\gamma\,, \nonumber \\ && 
E_{cut,\nu}=0.59  E_{cut,\gamma}\,.
\label{eq:params_nu}
\end{eqnarray}

We will use Eq.~\eqref{eq:flux_nu} and
Eq.~\eqref{eq:params_nu} for the calculation of the neutrino flux from
the different sources and to quantitatively evaluate the response of
Icecube. Following Ref.~\cite{GonzalezGarcia:2009jc} we will simulate the
detection at the level of the secondary muons. This allows us to use the
direct measurement of the muon energy to better characterize the expected
signal. 

Before describing in details the calculation of the muon events, we here want to 
comment on the difference between the parametrization of $\gamma$-ray that we use, 
Eq.~\eqref{eq:flux_nu}, and the one used in the literature, Eq.~\eqref{eq:Ngamma}. 
Indeed, we don't use a MonteCarlo simulation in our analysis to relate the gamma ray 
and neutrino fluxes, thus, we have to rely on the analytical approximation, described 
just above, that is based on a specific parametrization of the gamma ray flux. 
This could constitute a possible limitation of this kind of analytical study, but in 
the following we will describe how we have overcome this point. 
For each sources, we chose the parameters K, $\alpha_{\gamma}$ 
and $E_{cut,\gamma}$ in such a way to cover all the experimental data available. 
In particular, we have considered different scenarios when 
choosing the parametrization to use. 
With increasing $\alpha_\gamma$, we have spanned the low-energy spectra of each 
sources from lower to higher values of the spectra. On the other hand, with increasing $\alpha_\gamma$, 
the spectra will go from an harder to a softer spectra at high energies. 
Moreover, in choosing the different values of $E_{cut, \gamma}$, we have considered, for the 
first two sources and the lowest values of $\alpha_\gamma$, 
one scenario in which the spectrum is within the high energy Milagro 1$\sigma$ region up to 
roughly 35~TeV, and two other conservative scenarios in which the Milagro results 
at high energy could be relaxed and an harder spectrum is allowed. However, for these latter 
two cases, we have required that the spectrum is within the Milagro 1$\sigma$ band up to 20~TeV. 

\begin{figure}[!t]
\centering
\begin{tabular}{lcr}
\includegraphics[width=0.32\textwidth]{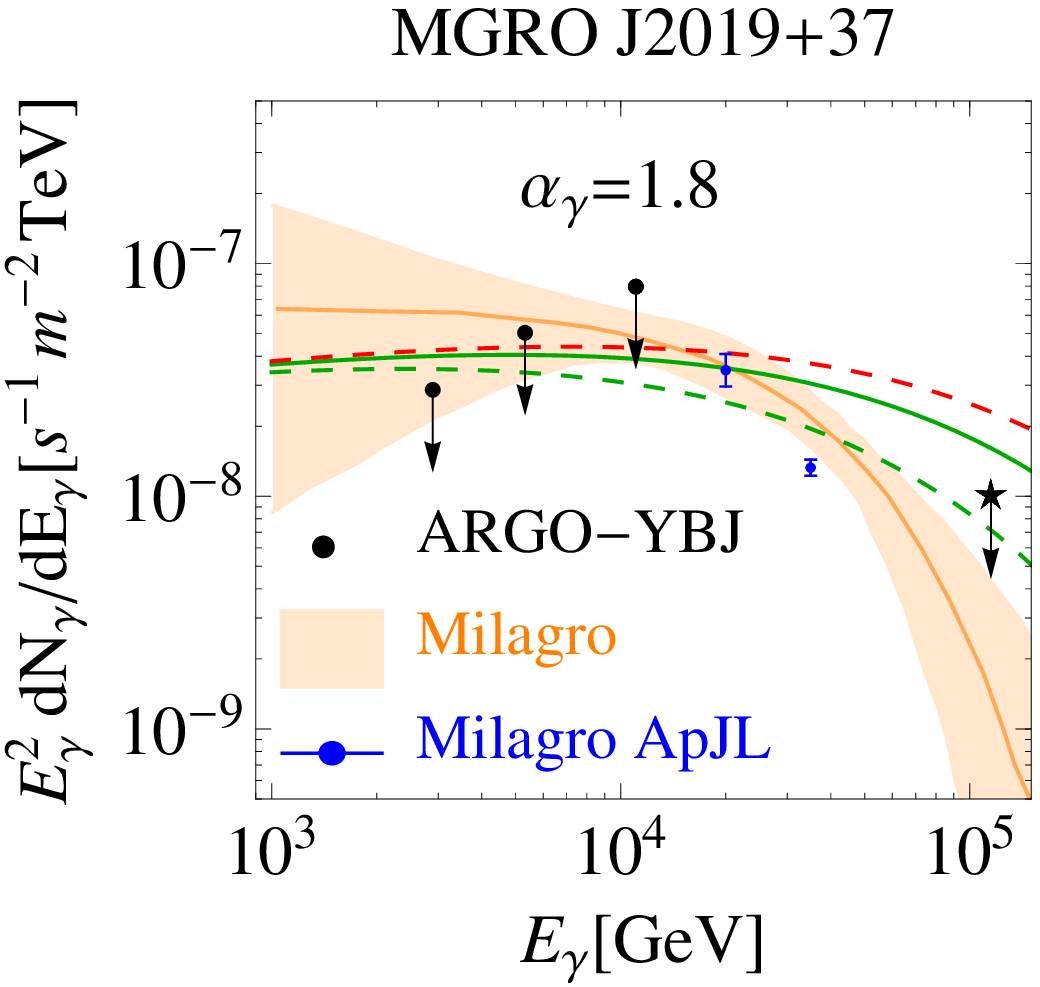} & 
\includegraphics[width=0.32\textwidth]{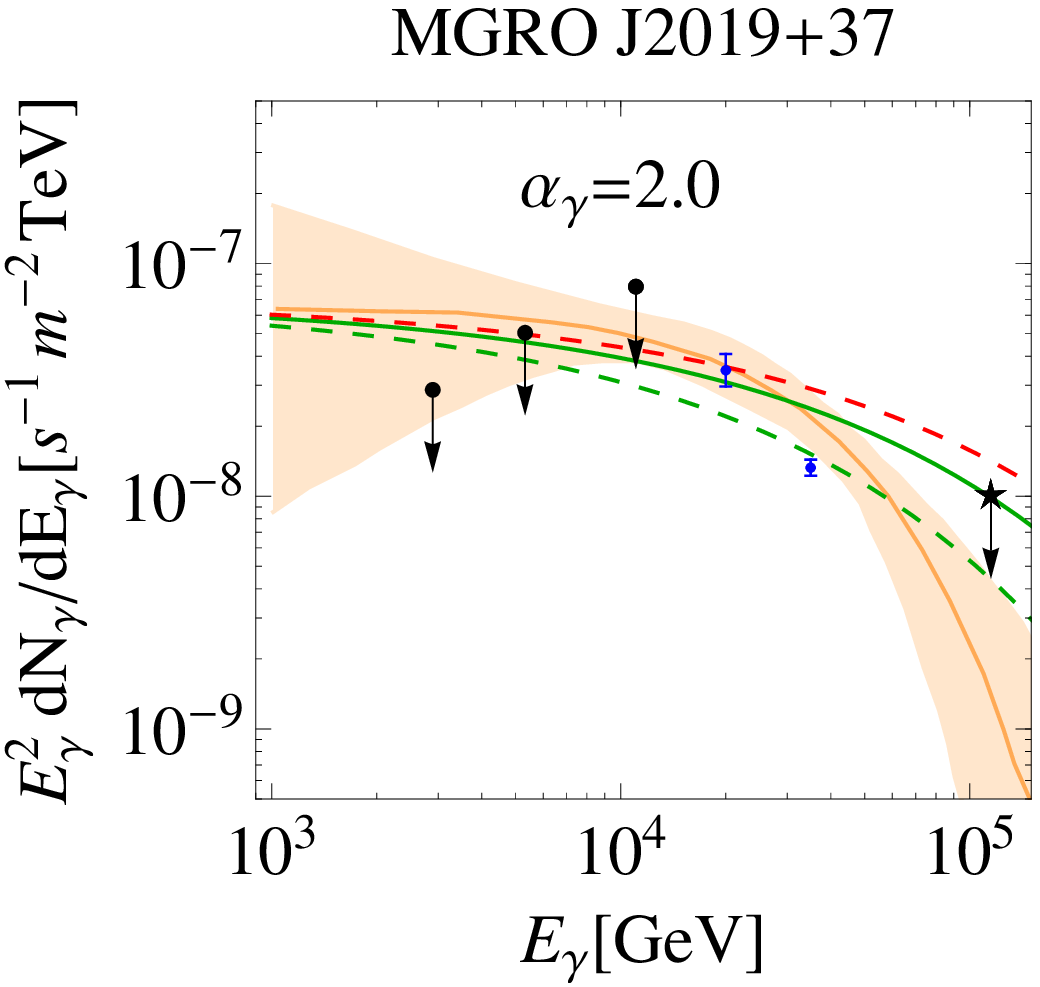} & 
\includegraphics[width=0.32\textwidth]{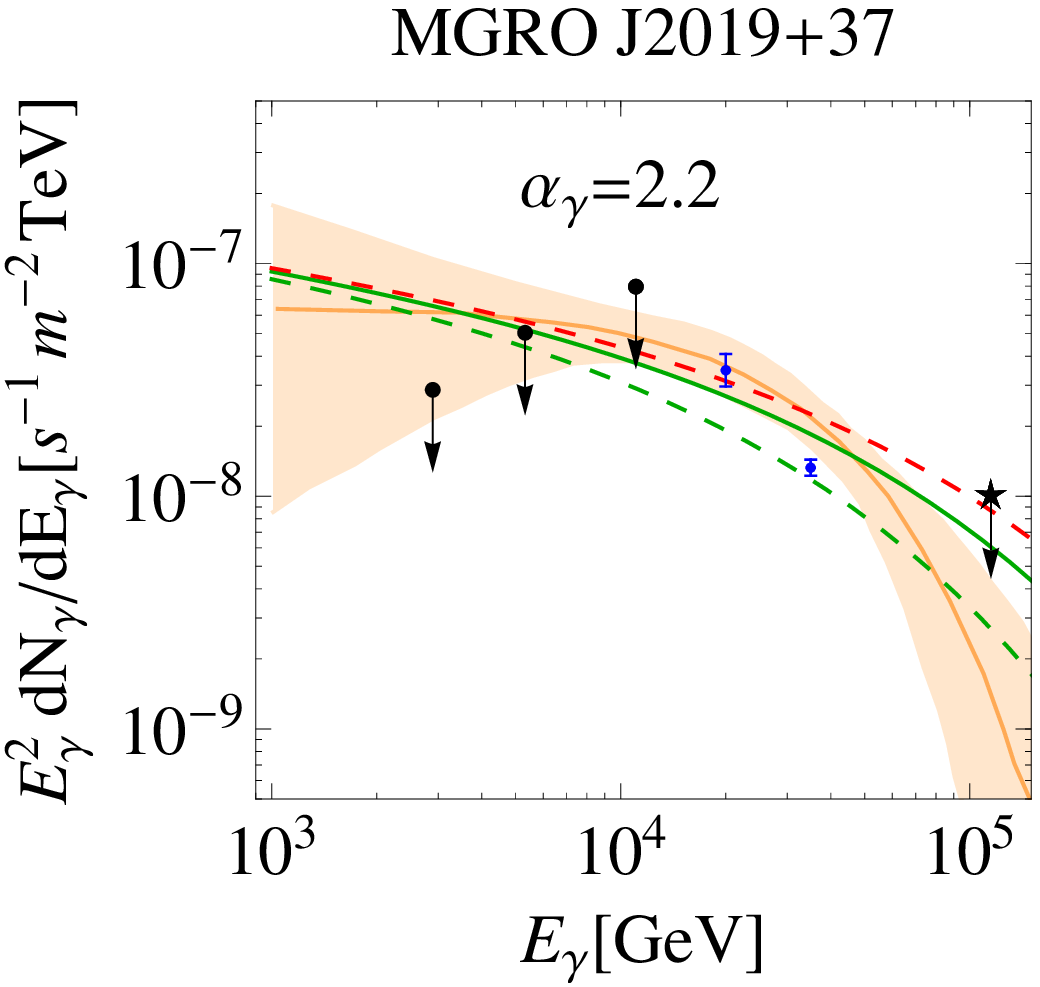} \\[10pt]
\includegraphics[width=0.32\textwidth]{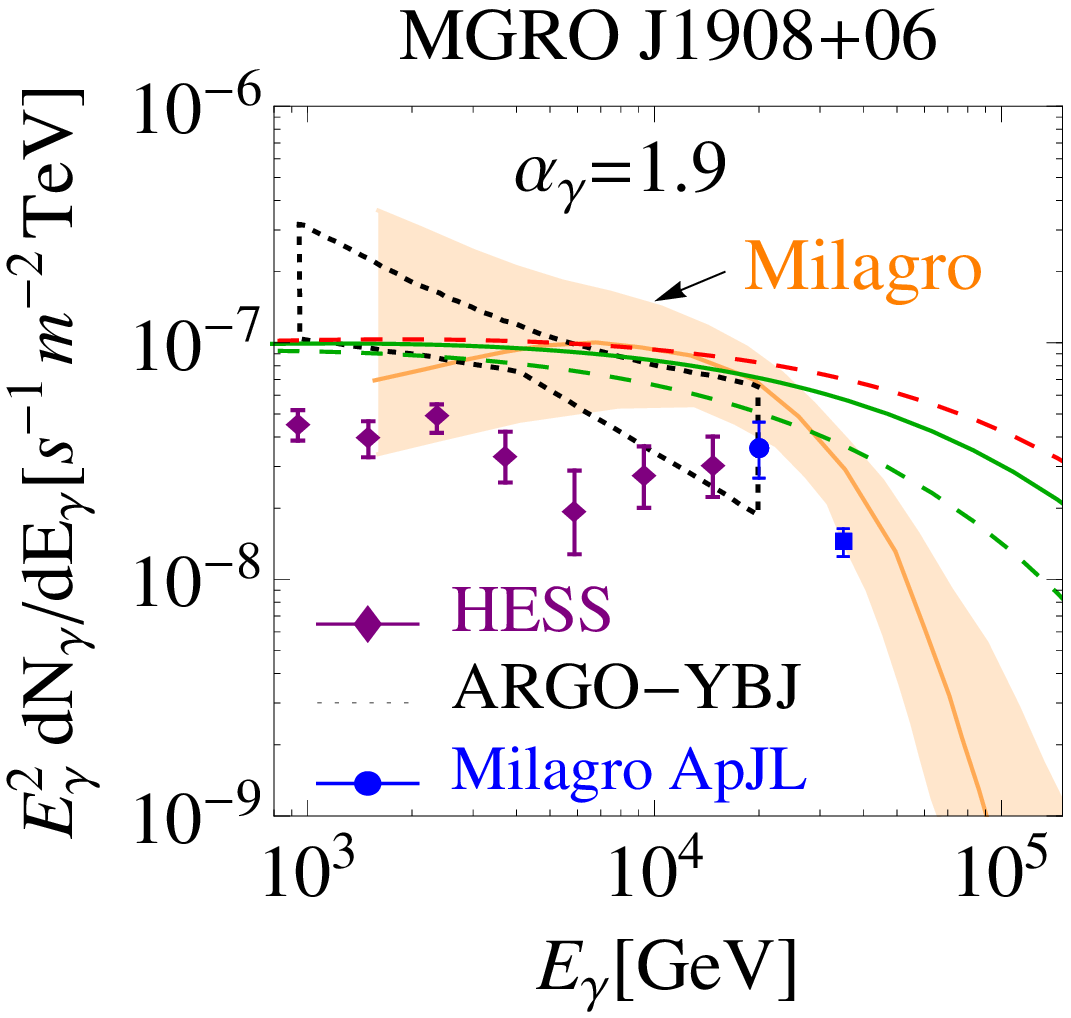} & 
\includegraphics[width=0.32\textwidth]{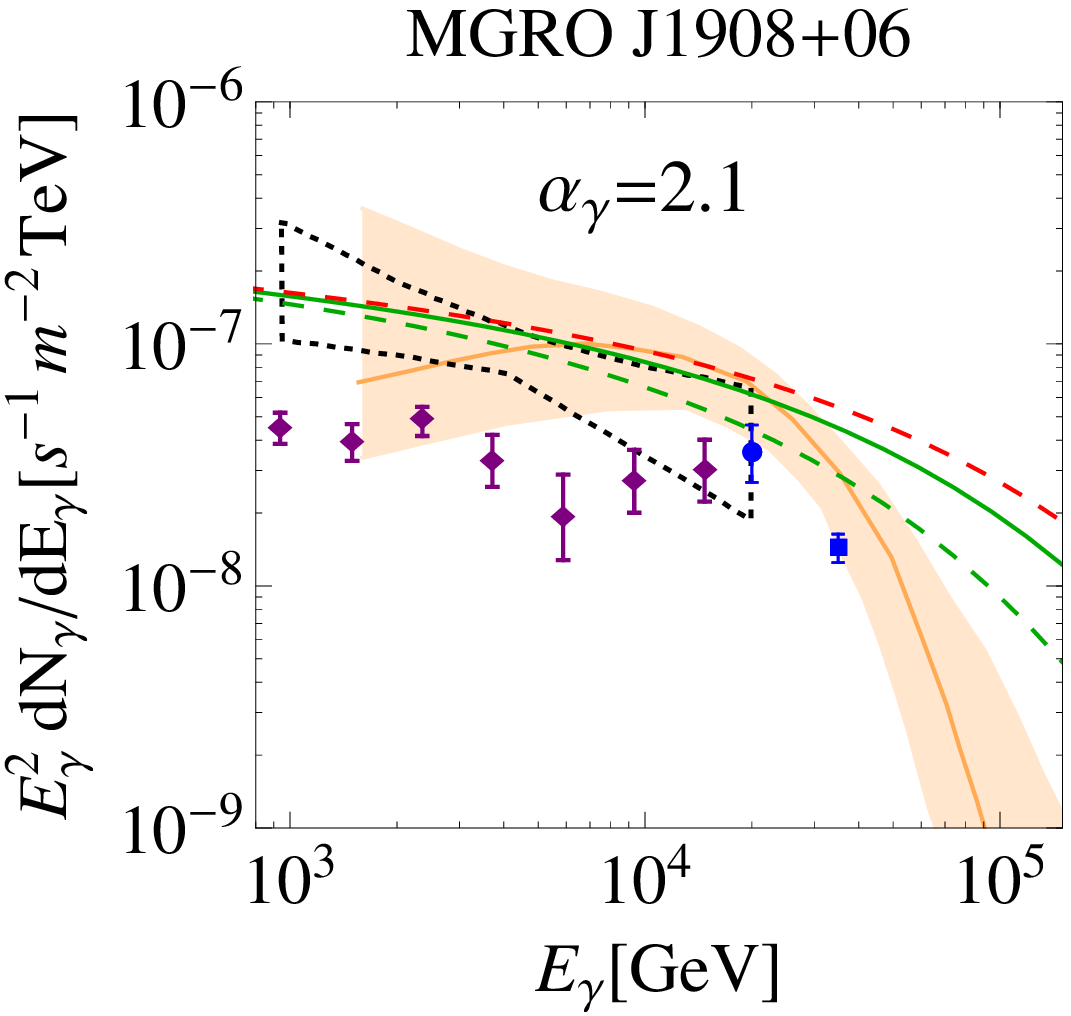} & 
\includegraphics[width=0.32\textwidth]{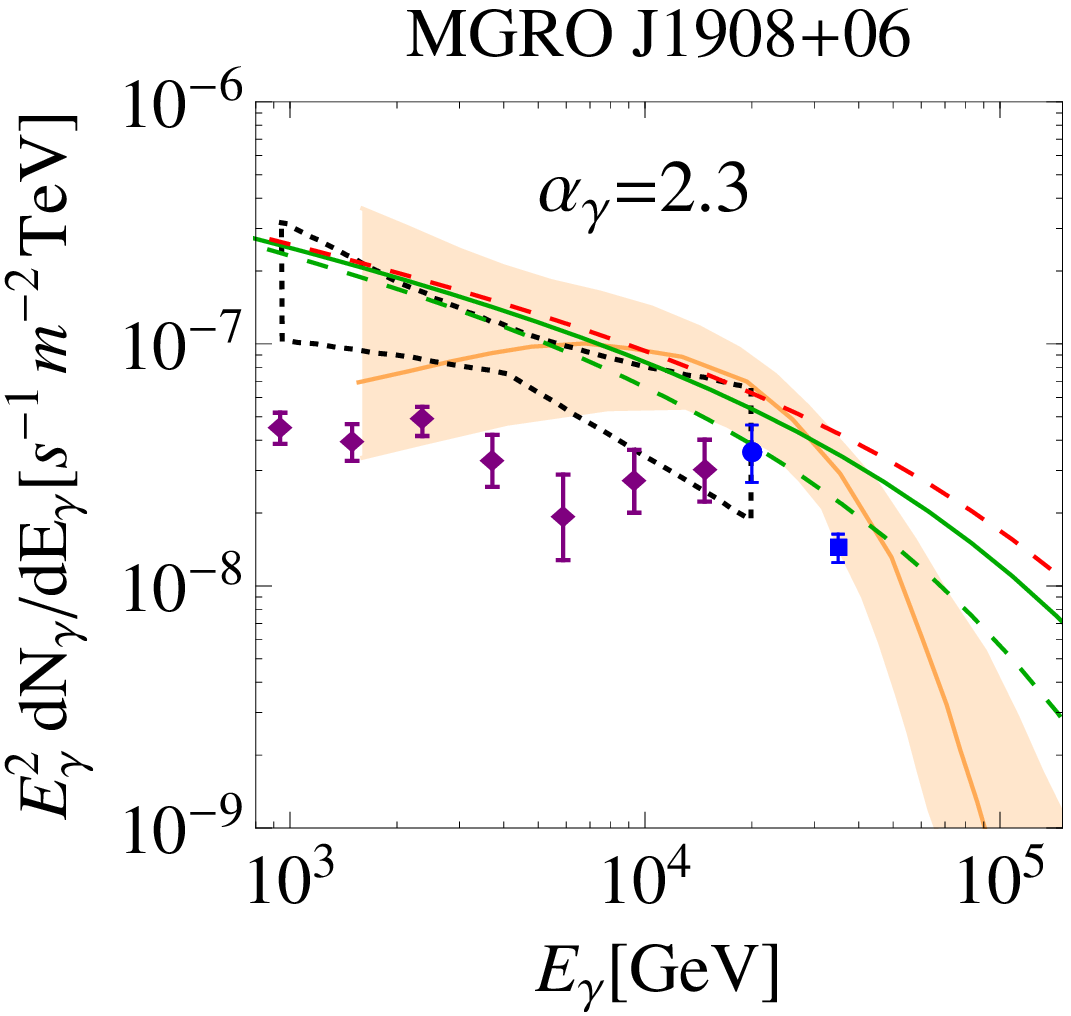} \\[10pt]
\includegraphics[width=0.32\textwidth]{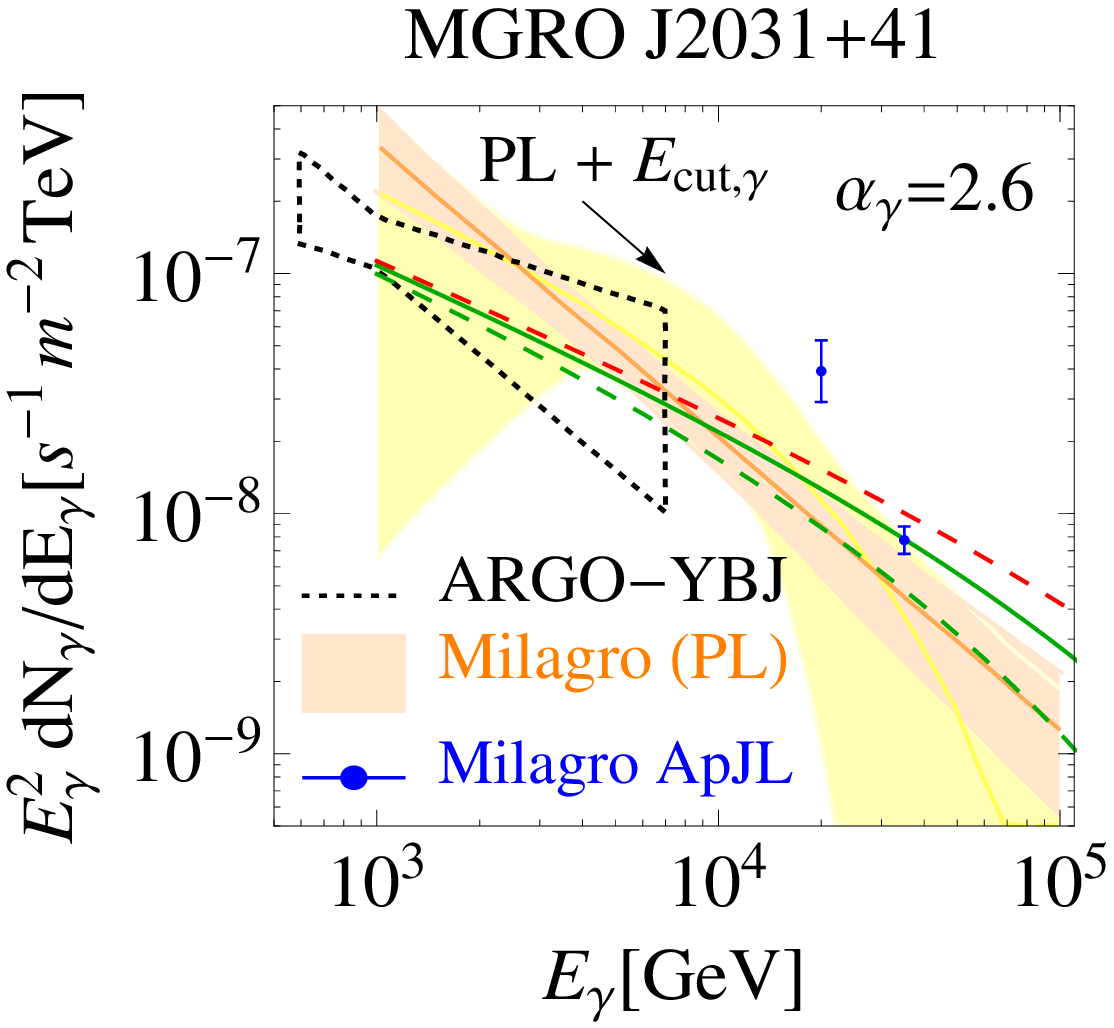} & 
\includegraphics[width=0.32\textwidth]{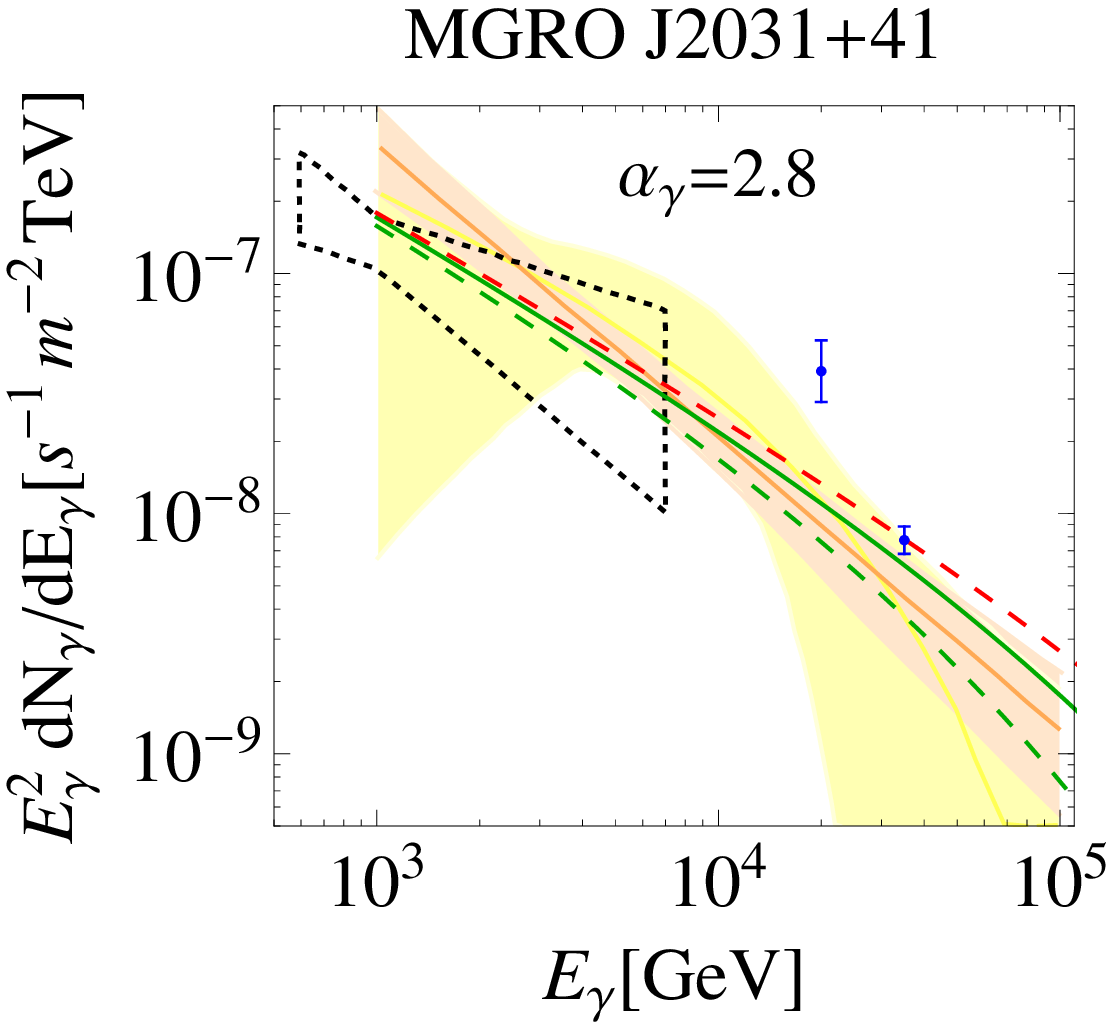} & 
\includegraphics[width=0.32\textwidth]{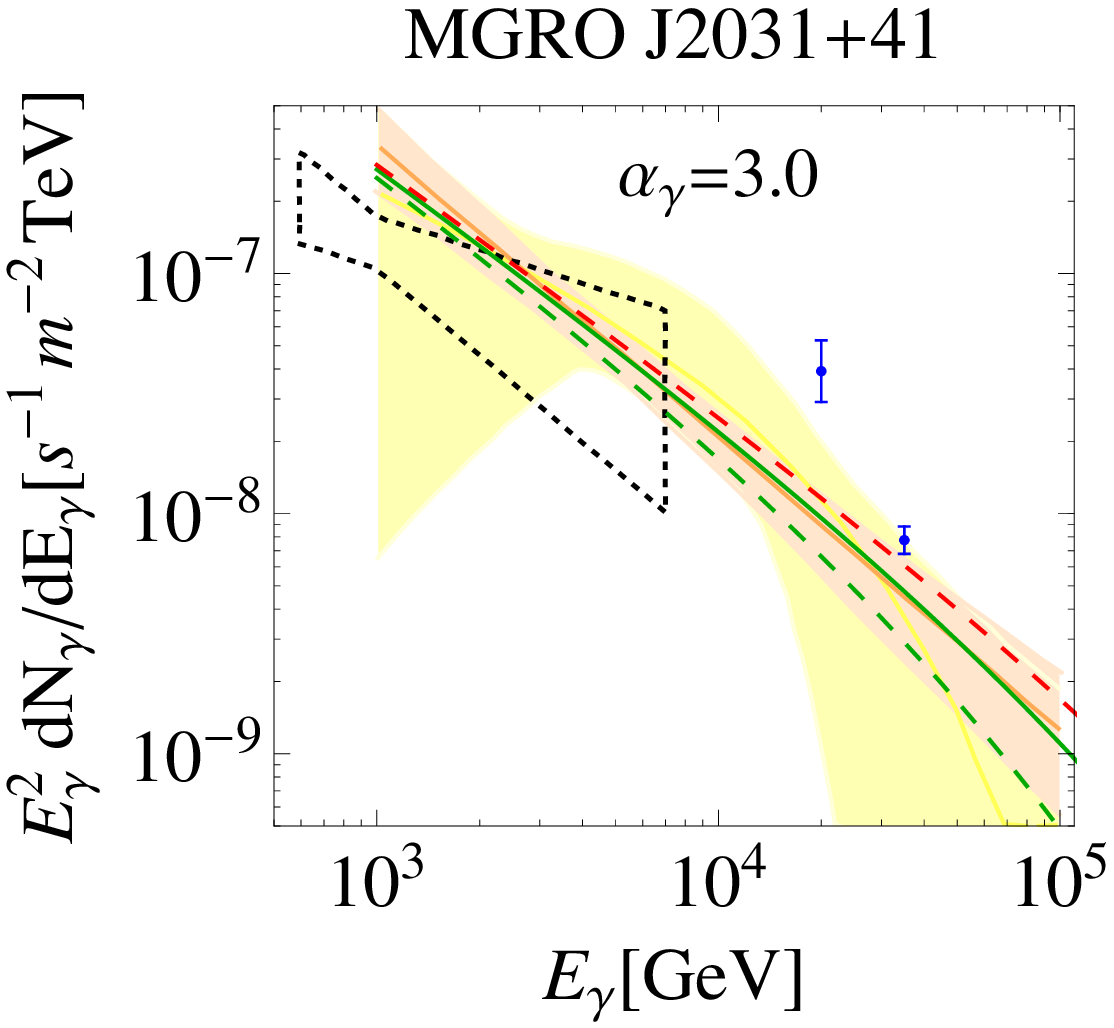}
\end{tabular}
\caption{\label{fig:sources_plot_1} 
Flux of $\gamma$-rays using our parametrization of
Eq.~\eqref{eq:Ngamma}.  The source flux has been normalized to the
values of K reported in Table~\ref{tab:sources_norm}.  The green
dashed, solid green and red dashed lines refer to the values of $E_{cut,\gamma}$ 
as reported in Table~\ref{tab:sources_norm}, from smallest to biggest values.  For MGRO J2019+37, the continuous orange
line is the best fit to the Milagro data, while the shaded orange area
represents the $1\sigma$ band~\cite{Abdo:2012jg}.  With blue lines, we
report also the previous flux measurements by Milagro (only statistical errors are reported) 
at 20 TeV and 35 TeV~\cite{Abdo:2007ad,Abdo:2009ku}. The 90\% CL upper limits from
ARGO-YBJ are shown in black~\cite{Bartoli:2012tj}, while the inferred CASA-MIA 
bound of~\cite{Beacom:2007yu} is reported with a black star. 
For MGRO~J1908+06, we show in purple the data by HESS~\cite{Aharonian:2009je} and in blue
the previous flux measurements by
Milagro~\cite{Abdo:2007ad,Abdo:2009ku}.  The dotted area shows the
ARGO-YBJ $1\sigma$ band~\cite{ARGO-YBJ:2012goa}, while the solid
orange line and the shaded orange area show the best fit and the
$1\sigma$ band by Milagro~\cite{Smith:2010yn}.  For MGRO~J2031+41,
the power law model is shown in orange and the power law model with
cut-off is shown in yellow~\cite{Abdo:2012jg}. The previous flux
measurements by Milagro  are 
shown in blue~\cite{Abdo:2007ad,Abdo:2009ku}. 
Note that for MGRO~J2031+41, we didn't report the measurements by
MAGIC~\cite{Albert:2008yk}, HEGRA~\cite{Aharonian:2005ex} and 
Whipple~\cite{Lang:2004bk}, see text for more details. 
}
\end{figure}

In Fig.~\ref{fig:sources_plot_1}, the flux of $\gamma$-rays are 
presented using our parametrization of
Eq.~\eqref{eq:Ngamma}. The values of K, $\alpha_\gamma$ and $E_{cut,\gamma}$ are 
the one reported in Table~\ref{tab:sources_norm}.  The green
dashed, solid green and red dashed lines refer to the values of $E_{cut,\gamma}$ 
as reported in Table~\ref{tab:sources_norm}, from smallest to biggest values. 
For the three sources, we reported in the plots the respective experimental data, that 
we will describe in details in the following. 
The continuous orange line is the best fit to the Milagro data, while the shaded orange area
represents the $1\sigma$ band~\cite{Abdo:2012jg}.  With blue lines, we
report also the previous flux measurements by Milagro at 20 TeV and 35
TeV~\cite{Abdo:2007ad,Abdo:2009ku}. For these two measurements only 
statistical errors are presented. For an estimation of the systematic errors, we 
refer to~\cite{Abdo:2007ad,Abdo:2009ku}. For MGRO J2019+37, the 90\% CL upper limits from
ARGO-YBJ~\cite{Bartoli:2012tj} are shown with black dots. 
We have also reported with a black star, the CASA-MIA bound at 100 TeV, as inferred 
in Ref.~\cite{Beacom:2007yu}. 
Note that the parametrization that we considered for $\alpha_\gamma=2.2$ and $E_{cut,\gamma}$=45~TeV 
is similar to the parametrization used in Ref.~\cite{Beacom:2007yu}. 

For MGRO~J1908+06, the dotted area shows the
ARGO-YBJ $1\sigma$ band~\cite{ARGO-YBJ:2012goa}, while the solid
orange line and the shaded orange area show the best fit and the
$1\sigma$ band by Milagro~\cite{Smith:2010yn}. The blue points are the 
previous flux measurements reported by Milagro~\cite{Abdo:2007ad,Abdo:2009ku}. 
We have scanned with the different $\alpha_\gamma$ the different limiting 
cases of compatibility with the ARGO-YBJ $1\sigma$ band. 
We show in purple the data by HESS~\cite{Aharonian:2009je}, 
that are systematically lower than the other data. 
The discrepancy between these two different data sets is between 2$\sigma$ and 
3$\sigma$~\cite{Smith:2010yn}. This could be, in principle, due also to statistical fluctuation. 
However, we want to point out that the source MGRO~J1908+06 is not point-like and, for this reason, 
the Milagro detector, that has a worse angular resolution respect to HESS, observes a much higher flux. 
The HESS detector, indeed, would just detect the flux from the source core. 
Moreover, the HESS collaboration finds a spectrum with no evidence of a cut off, but their energy reach is limited to $<20$~TeV. 
Note, finally, that any loss in sensitivity because this source is near the horizon is taken into account in our 
calculation by the effective area.  

For MGRO~J2031+41, the power law model is shown in orange and the power law model with
cut-off is shown in yellow~\cite{Abdo:2012jg}. The dotted area shows the 
ARGO-YBJ $1\sigma$ band~\cite{ARGO-YBJ:2012goa}, while the blue points are the 
previous flux measurements reported by Milagro~\cite{Abdo:2007ad,Abdo:2009ku}. 
Note that, we didn't report the measurements by
MAGIC~\cite{Albert:2008yk} and HEGRA~\cite{Aharonian:2005ex}.  These
measurements are mutually consistent, but they disagree with the best
fit and $1\sigma$ region obtained by Milagro~\cite{Abdo:2012jg},
considering both a power-law and a power-law with energy cut-off. 
In particular, at low energies, the flux measured by air Cherenkov telescopes (ACTs) 
is much smaller than the flux measured by Milagro (it accounts for just the 
3\% of the Milagro flux) and is much harder than 
the power law best-fit given by Milagro. 
As described in details in Ref.~\cite{Abdo:2012jg}, the discrepancy between 
the experimental data at low energies could be explained by different reasons. 
First of all, the angular regions considered by the experiments are different, since 
the angular resolution of ACTs experiments is better than the one of 
Milagro. For this reason, the Milagro detector measures photons coming
from a larger region around the actual position of the source respect to the ACTs measurements. 
Note, however, that the flux at 0.6~TeV was measured also by Whipple, see
Ref.~\cite{Lang:2004bk} for more details, and this measurements agrees well with the 
extrapolation of the Milagro result at lower energies. The same is true for ARGO-YBJ, which 
has an angular resolution similar to Milagro. 
Second, the way the background is computed is also different. 
Indeed, the source MGRO J2032+41 has an extension slightly
larger than the HEGRA and MAGIC angular resolution and is surrounded by an extended emission. 
Therefore, Milagro, Whipple and ARGO-YBJ are not able to disentangle
the extended emission from the central source and
observe a higher flux, while MAGIC and HEGRA consider the extended
emission as a background. 
Note that with the different values of $\alpha_\gamma$ that we considered in this paper, we have 
spectra more in agreement with the ARGO-YBJ data and spectra which follow more the Milagro spectrum. 
According to Ref.~\cite{Fox:2013oza} the Milagro spectrum for this source yields neutrino predictions lower by roughly 
a factor of two respect to the ARGO-YBJ.

We moreover want here to comment on some recent results by Fermi. In particular, the pulsar PSR J1907+0602 was detected to 
be within the extension of the Milagro source MGRO J1908+06~\cite{Abdo:2010ht}. Thus, this TeV source 
can be considered as the pulsar wind nebula of PSR J1907+0602. 
In particular, using Fermi data, 2$\sigma$ upper limits on the source flux between $10^{-1}$ and $10^2$~GeV were found, 
see Fig.~4 of Ref.~\cite{Abdo:2010ht}. 
We decided not to introduce these data explicitly in the analysis, because 
since these data span different energy scales, 
it might be difficult to find a common parametrization, considering our Eq.~\eqref{eq:Ngamma}. 
Since we are going to consider only muons with energies above 10$^3$ GeV, we decided to find parametrizations that could 
describe the high energy part of the $\gamma$-ray fluxes and eventually also the energy cut-off reported by Milagro. 
Note also that other Fermi data for the Cygnus region, have been used recently in the Ref.~\cite{Tchernin:2013wfa}. 
In particular, the authors found that considering a 4 degrees radius region, the two sources MGRO J2019+37 and MGRO J2019+41 
are within the Fermi field of view and the sum of the Milagro 
1$\sigma$ region for the two sources can be considered in agreement with the Fermi data at low energies, 
see Fig.~4 of ~\cite{Tchernin:2013wfa} for more details.

\begin{figure}
\centering
\begin{tabular}{lcr}
\includegraphics[width=0.32\textwidth]{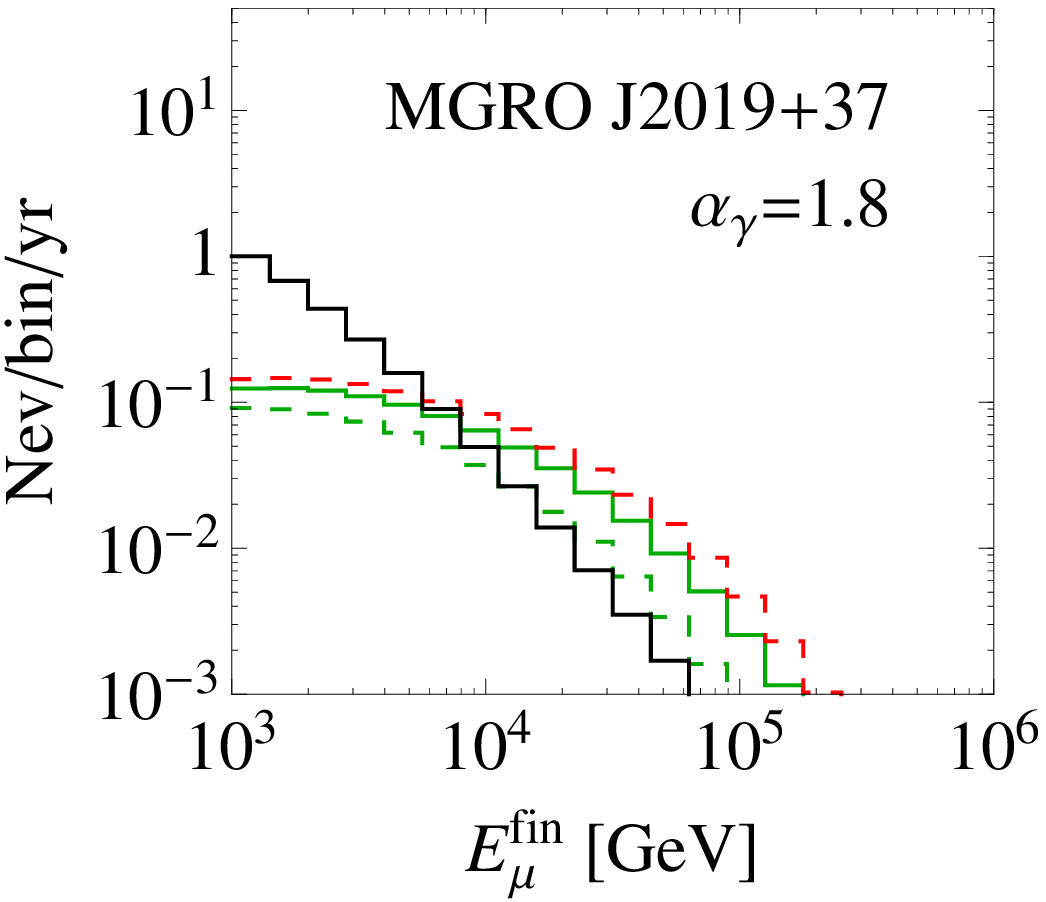} & 
\includegraphics[width=0.32\textwidth]{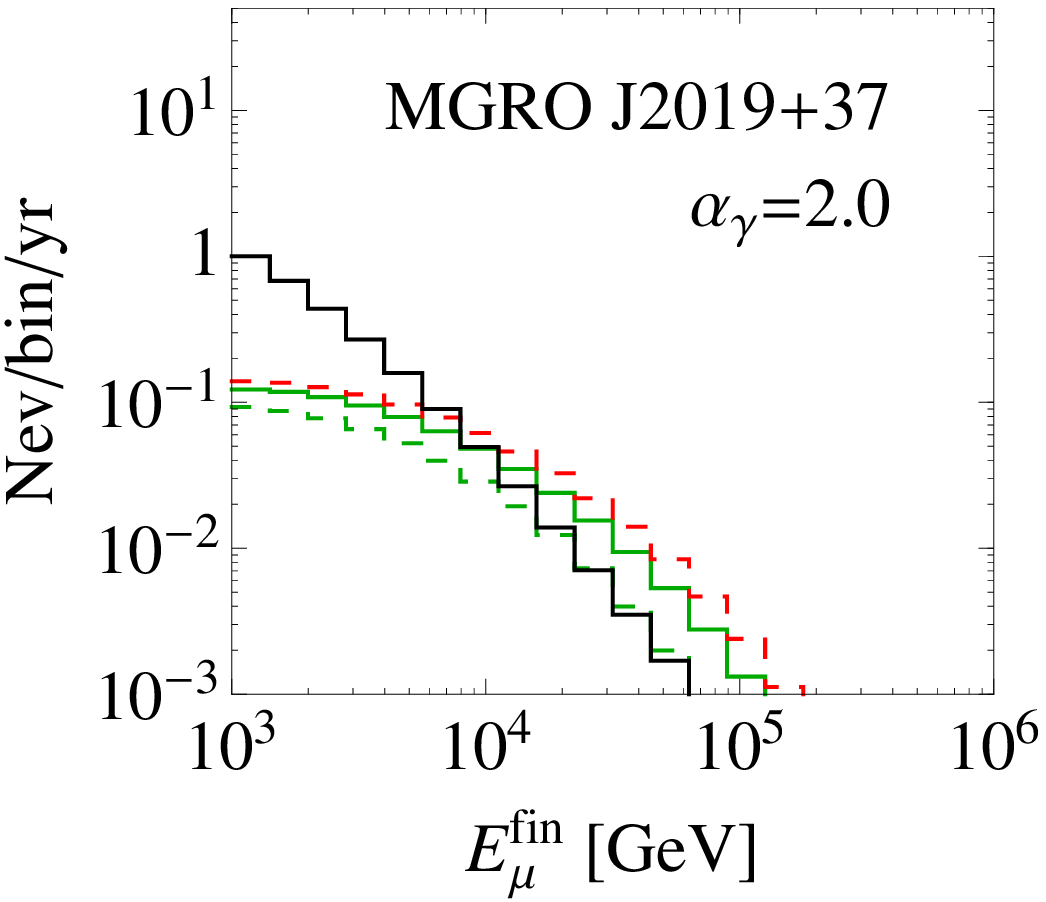} &
\includegraphics[width=0.32\textwidth]{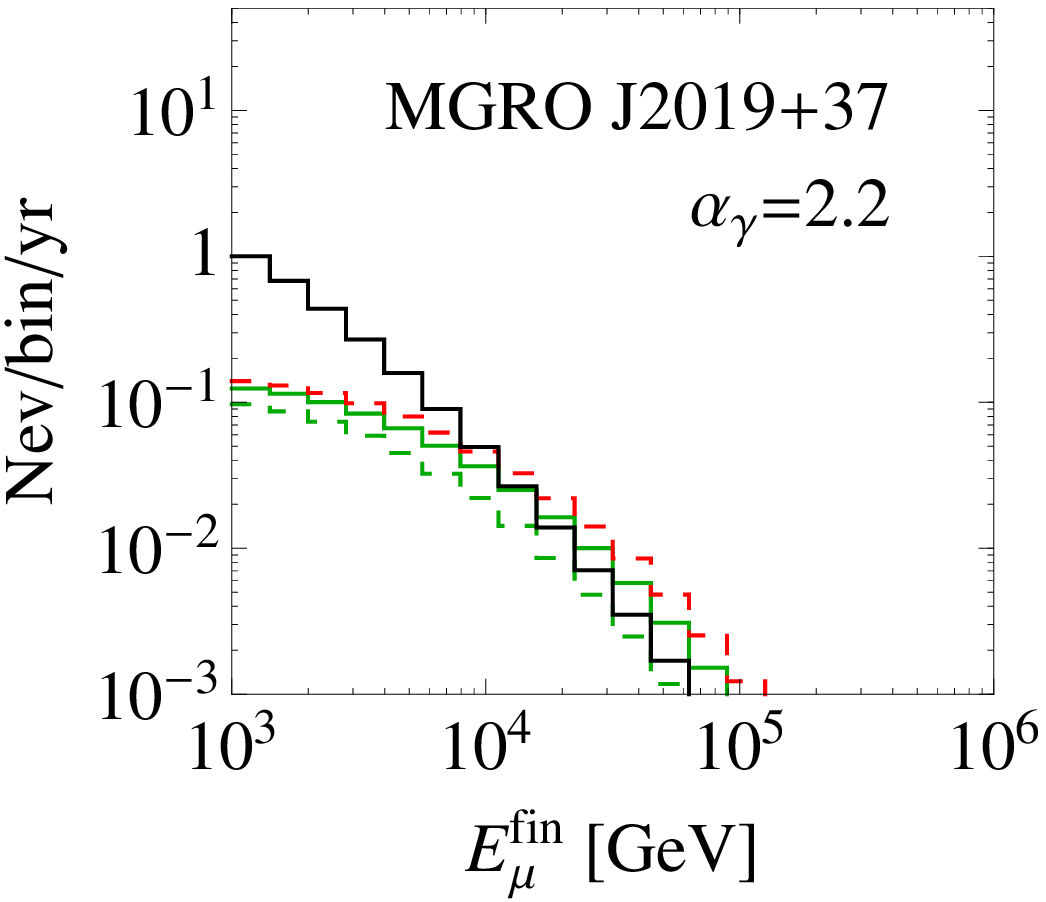} \\
\includegraphics[width=0.32\textwidth]{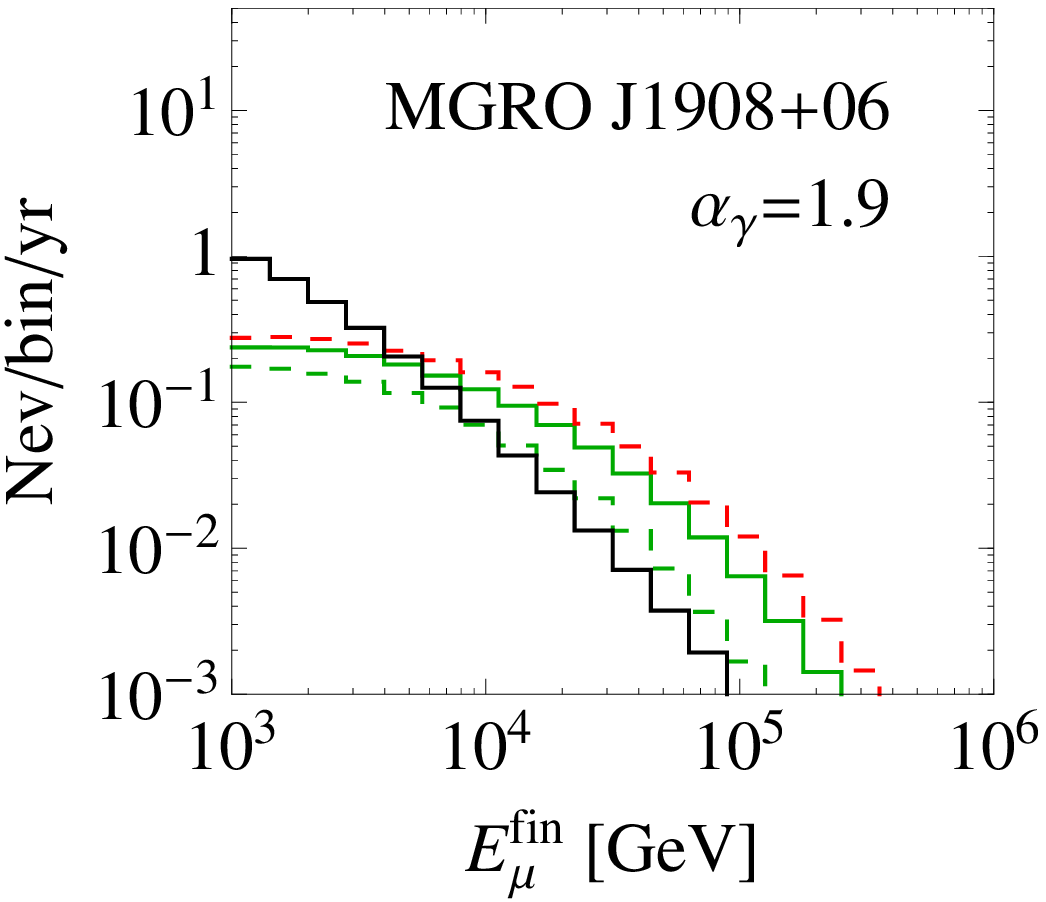} &
\includegraphics[width=0.32\textwidth]{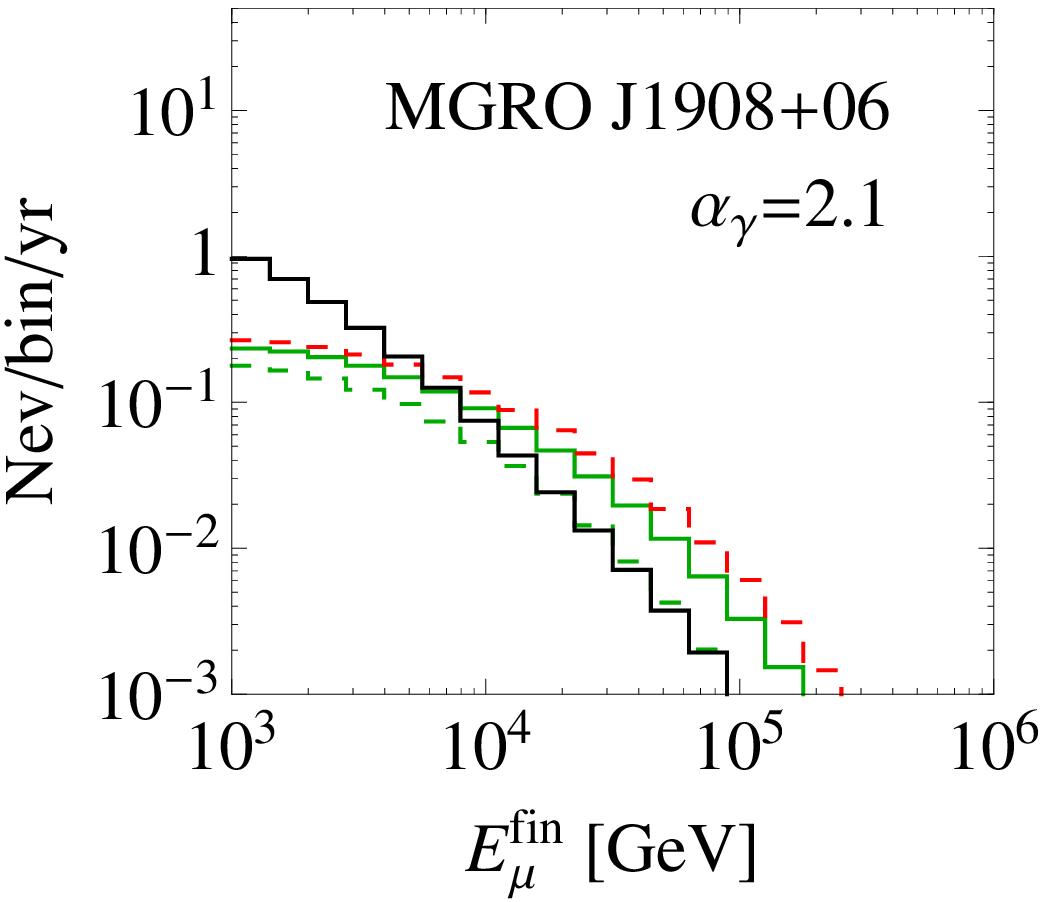} &
\includegraphics[width=0.32\textwidth]{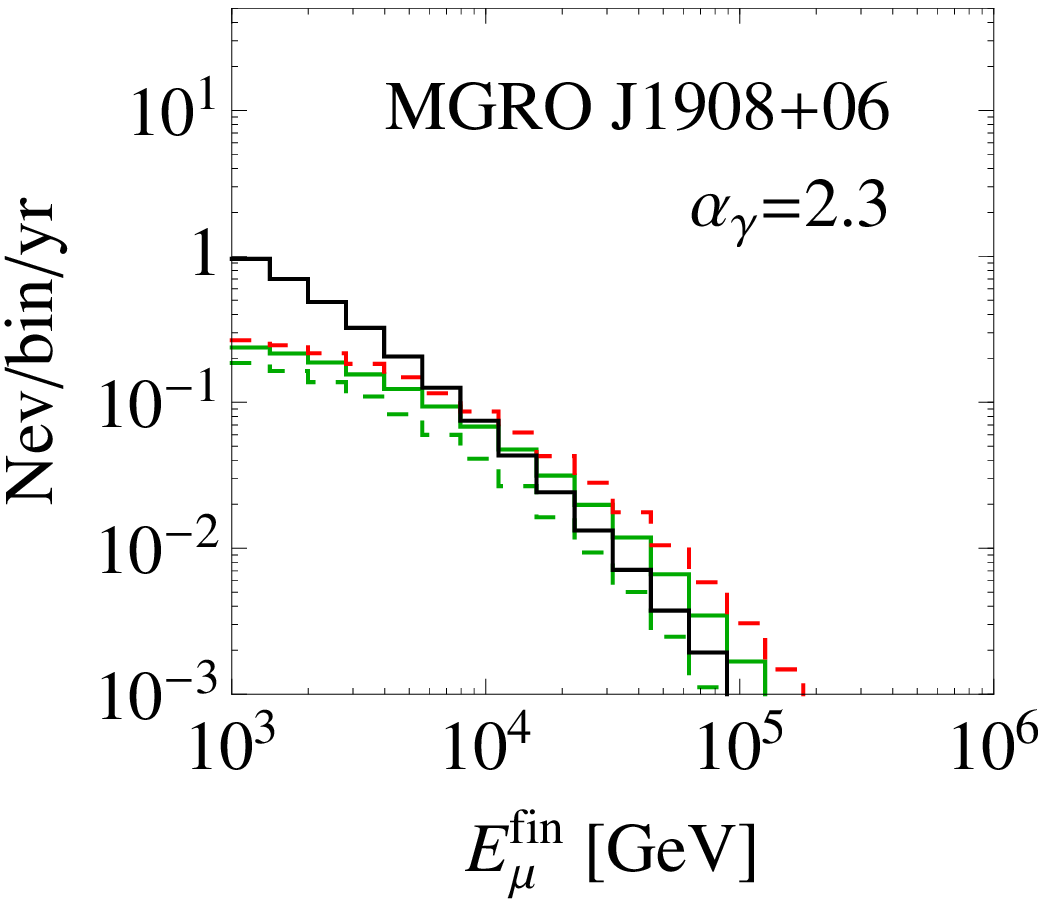} \\
\includegraphics[width=0.32\textwidth]{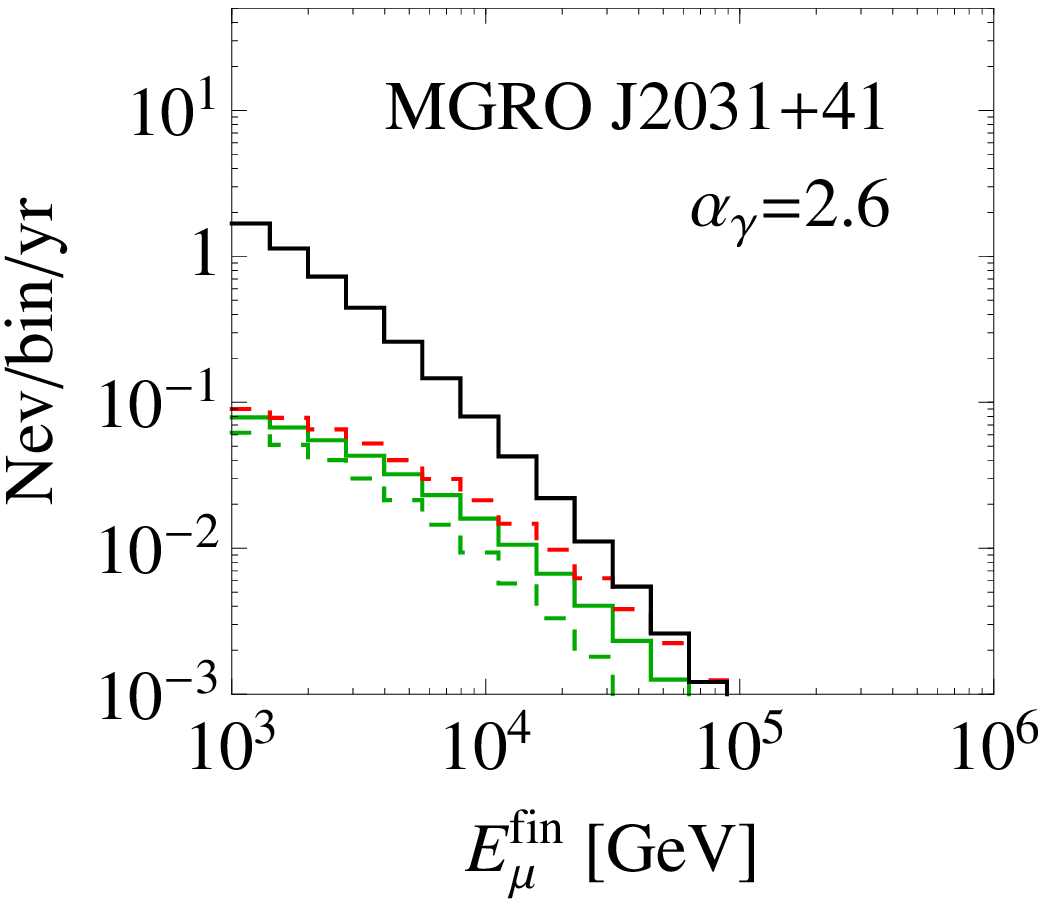} &
\includegraphics[width=0.32\textwidth]{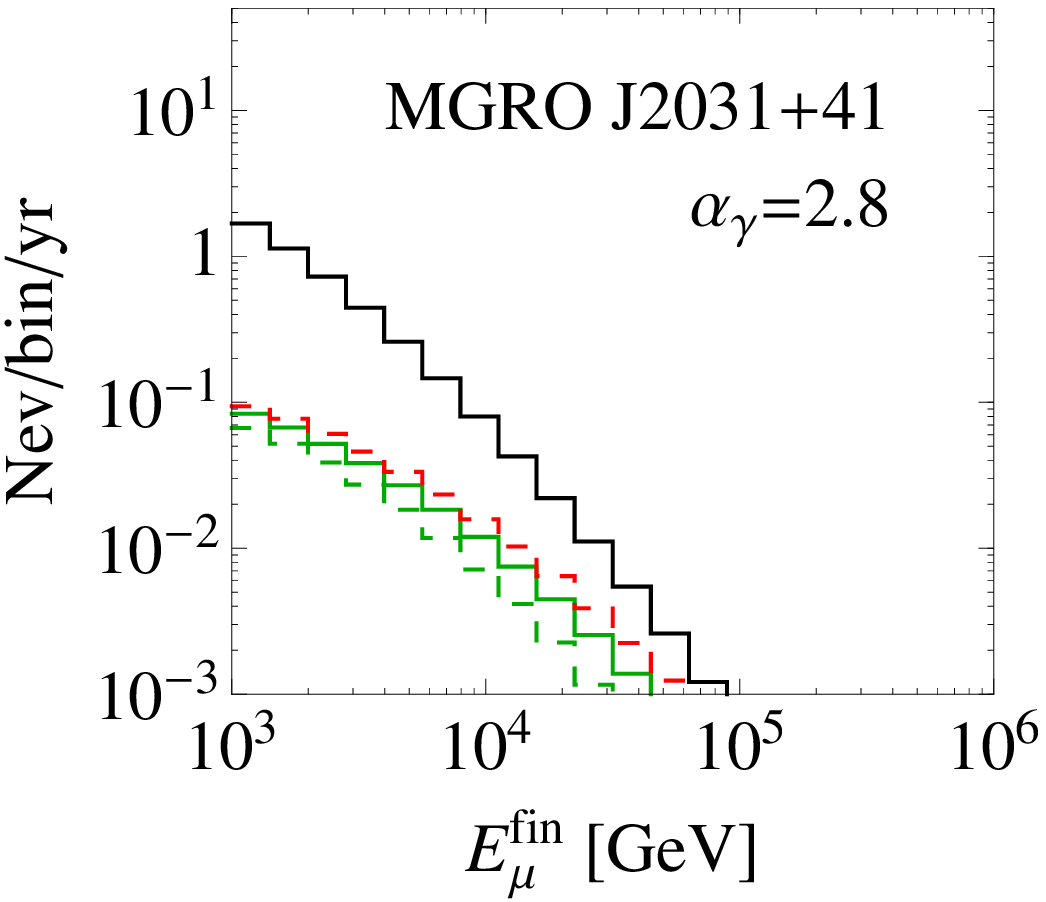} &
\includegraphics[width=0.32\textwidth]{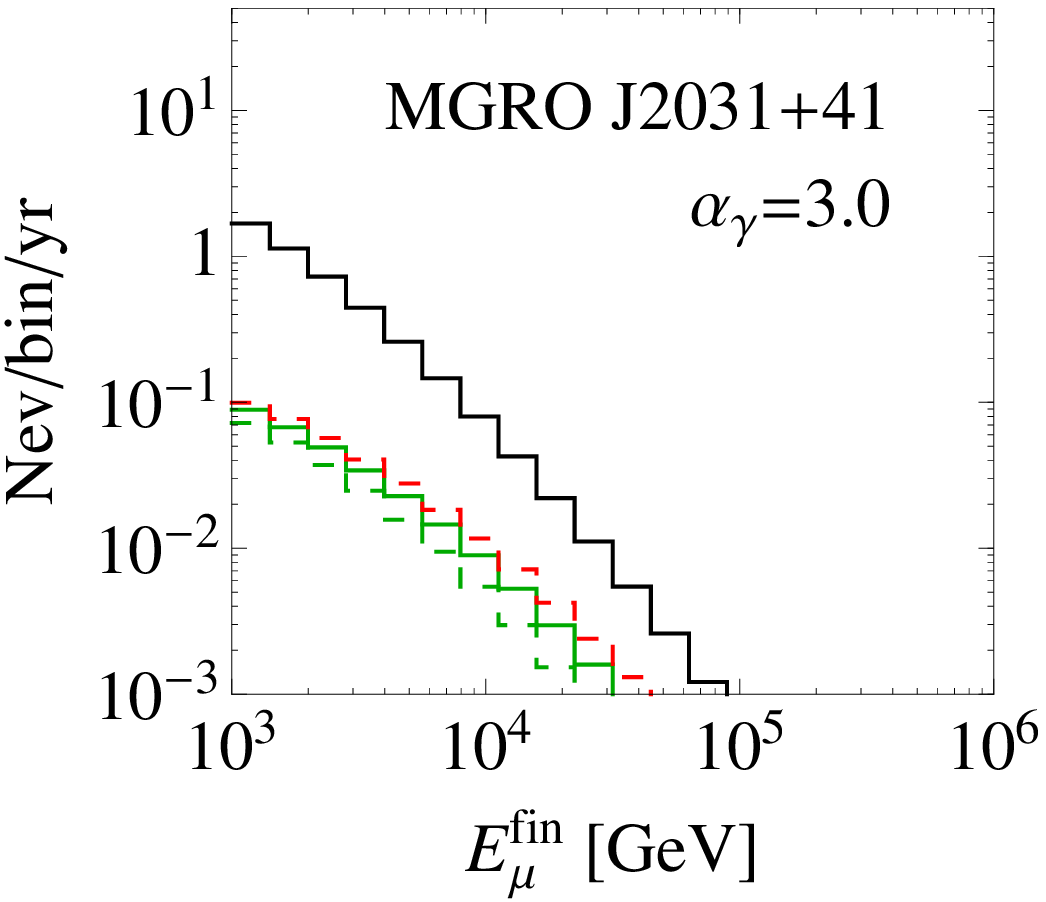} 
\end{tabular}
\caption{\label{fig:events_sources} Number of events for the three
Milagro sources with different values of $\alpha_\gamma$. The green
dashed, solid green and red dashed lines refer to the values of $E_{cut,\gamma}$ 
as reported in Table~\ref{tab:sources_norm}, from smallest to biggest values. The source fluxes have been normalized as
reported in Table~\ref{tab:sources_norm}.  The black lines represent
the atmospheric neutrino background, integrated over a solid angle
equal to $\Omega = \pi (1.6~\sigma)^2$, with $\sigma= 0.9^\circ$ for
MGRO J2019+37 and $0.6^\circ$ for MGRO J1908+06 and $1.2^\circ$ for
MGRO J2031+41. A correction for the size of the source has been
introduced (factor 72\%).}
\end{figure}

\begin{table}[!t]
\centering
\begin{tabular}{|l || c | c | c | c |}
\hline
Source & 
K~[TeV$^{-1}$ cm$^{-2}$ s$^{-1}$] & 
$E_{norm,i}$~[TeV] & 
$\alpha_{\gamma,i}$ & $E_{cut,\gamma, i}$~[TeV] \\ [1ex]
\hline
MGRO J2019+37 & 7$^{+5}_{-2}$ $\times$ 10$^{-14}$ & 10 
& $2.0^{+0.5}_{-1.0}$ & $29^{+50}_{-16}$ \\[6pt]
MGRO J1908+06 & 6.1$^{+1.4}_{-1.4}$ $\times$ 10$^{-13}$ & 4 
& $2.54^{+0.36}_{-0.36}$ & -- \\[6pt]
MGRO J2031+41$^{(a)}$ & 2.1$^{+0.6}_{-0.6}$ $\times$ 10$^{-14}$ & 10 
& $3.22^{+0.23}_{-0.18}$ & -- \\[6pt]
MGRO J2031+41$^{(b)}$ & 5$^{+157}_{-3}$ $\times$ 10$^{-14}$ & 10 
& $2.7^{+0.7}_{-3.3}$ & $21^{+\infty}_{-18}$ \\\hline
\end{tabular}
\caption{Best-fit values for MGRO J2019+37, MGRO J1908+06 and MGRO
J2031+41, as reported in
Refs.~\cite{ARGO-YBJ:2012goa,Abdo:2012jg} from the ARGO-YBJ and the Milagro 
experiments. For the MGRO J2031+41
source, we report the values for a power law fit (a) and a power law
with cut-off fit (b).}
\label{tab:sources}
\end{table}

\begin{table}[t]
\centering
\begin{tabular}{|l || c | c | c | c |}
\hline
Source & 
K~[TeV$^{-1}$ cm$^{-2}$ s$^{-1}$] & $E_{norm}$~[TeV] &
$ \alpha_{\gamma,i}$ & $E_{cut,\gamma,i}~[\rm TeV]$ \\ [1ex]
\hline
MGRO J2019+37 & 7 $\times$ 10$^{-14}$ & 10 & $1.8, 2.0, 2.2$ & $15, 30, 45$ \\ \hline 
MGRO J1908+06 & 1.5 $\times$ 10$^{-13}$ & 10 & $1.9, 2.1, 2.3$ & $15, 30, 45$ \\ \hline 
MGRO J2031+41 & 3 $\times$ 10$^{-14}$ & 10 & $2.6, 2.8, 3.0$ & $30, 100, 300$ \\ \hline 
\end{tabular}
\caption{
Values of K, $E_{norm}$, $\alpha_{\gamma,i}$ and
$E_{cut,\gamma,i}$ that we used in our analysis. We refer to
Eq.~\eqref{eq:Ngamma} and Eq.~\eqref{eq:kgamma} for the definition of
the variables.}
\label{tab:sources_norm}
\end{table}

\subsection{\label{sec:events} Events}

The number of events detected by IceCube from a source at zenith angle
$\theta_Z$ can be written
as~\cite{GonzalezGarcia:2005xw,GonzalezGarcia:2009jc}
\begin{eqnarray}
N_{ev}= t\times N_T \, \int dE_\nu 
dE_\mu^0 dE_\mu^{fin}
&& \Big[
\frac{dN_\nu(E_\nu)}{dE_\nu} 
\times Att_\nu (E_\nu,\theta_Z)
\times
\frac{d\sigma_\nu(E_{\nu},E_{\mu}^{0})}{dE_{\mu}^{0}} \nonumber \\
&& 
\!\!\!\!\!\!\!\!\!\!\!\!\!\!\!\!\!\!\!\!\!\!\!\!
+\frac{dN_{\bar\nu}(E_\nu)}{dE_\nu}\times
Att_{\bar\nu} (E_\nu,\theta_Z)
\times \frac{d\sigma_{\bar\nu}(E_{\nu},E_{\mu}^{0})}{dE_{\mu}^{0}}
\Big] \nonumber \\
&&
\!\!\!\!\!\!\!\!\!\!\!\!\!\!\!\!\!\!\!\!\!\!\!\!
\!\!\!\!\!\!\!\!\!\!\!\!\!\!\!\!\!\!\!\!\!\!\!\!
\times RR(E_{\mu}^{0},E_{\mu}^{fin}) \times 
A_\mu^{eff}(E_\mu^{fin},\theta_Z)\,, 
\label{eq:nevmus}
\end{eqnarray}
where we have summed over neutrinos and antineutrinos
contributions. The neutrinos from the astrophysical sources or the
atmospheric background interact with a cross section
$\frac{d\sigma_\nu(E_{\nu},E_{\mu}^{0})}{dE_{\mu}^{0}}$ to yield a
secondary muon of energy $E_{\mu}^{0}$.  For the differential deep
inelastic cross section $d\sigma_\nu(E_{\nu},E_{\mu}^{0})/
dE_{\mu}^{0}$, we use the full expression (without any average
inelasticity approximation) obtained with the CT10 NNLO
PDFs~\cite{Gao:2013xoa} with $\alpha_s=0.118$.  For  $x\leq 10^{-5}$
the PDF's are extrapolated using the 
double-log-approximation~\cite{Gribov:1984tu} 
following Refs.~\cite{Quigg:1986mb,Reno:1987zf,Gandhi:1995tf}.
The muon propagation is taken into account by
$RR(E_{\mu}^{0},E_{\mu}^{fin})$, that describes the probability of a muon
produced with initial energy $E_\mu^0$  arriving at the detector with energy
$E_\mu^{\rm fin}$ after taking into account energy 
losses due to ionization, bremsstrahlung, $e^+e^-$ pair production and nuclear 
interactions~\cite{Lipari:1991ut}. We have used the continuous approximation 
for the energy loss. $N_T$ is
the target density of the material surrounding the detector, while 
$Att_{\nu(\bar\nu)}(E_\nu,\theta_Z)$ is a factor which accounts for
the attenuation of the flux due to neutrino (antineutrino) propagation
in the Earth: 
\begin{equation}
Att_{\nu(\bar\nu)}(E_\nu,\theta_Z)=
\exp[-X(\theta_Z)(\sigma_{\rm NC}(E)+\sigma_{\rm CC}(E))] \, ,
\label{eq:fluxapp}
\end{equation}
where $X(\theta_Z)$ is the column density of the Earth assuming the
matter density profile of the Preliminary Reference Earth
Model~\cite{Dziewonski:1981xy}. Finally 
the detector
performance is described by its effective area for detecting muons
$A_\mu^{eff}(E_\mu^{fin},\theta_Z)$ and the exposure time $t$. 
For the functional form of the effective area we use the one reported in 
Ref.~\cite{GonzalezGarcia:2009jc}. 

We use the Honda flux~\cite{Honda:2011nf} to calculate the background
of atmospheric neutrinos at the zenith angle corresponding to the
source. We extrapolate this flux to higher energies to match the one
from Volkova~\cite{Volkova:1980sw}, that is known to describe the
AMANDA data in the energy range of interest here.  At high energy, the
contribution of prompt neutrinos from charm decay cannot be
neglected. We consider in this paper the model of Thunman {\sl et al}
(TIG)~\cite{Gondolo:1995fq}. We integrate the atmospheric background
over a solid angle $\Omega=\pi (1.6 \sigma)^2$ around the direction of
the source. The angle $\sigma$ combines the effects of the angular
resolution of the detector and the size of the source. For the size of
the sources $\sigma_{\rm ext}$, we refer to
Table~\ref{tab:sources_position}, while the angular resolution of
IceCube $\sigma_{\rm IC}$ at these high energies is around
$0.5^\circ$. The angle $\sigma$ is then given by $\sqrt{\sigma_{\rm
ext}^2 + \sigma_{\rm IC}^2}$.  Assuming gaussianity, roughly 72\% of
the flux of the source is contained within this angular
bin~\cite{Alexandreas:1992ek}. We will bin the events in the measured
muon energy $E_\mu^{fin}$ assuming an energy resolution of 15\% in
$\log(E_\mu^{fin})$ for $E_\mu^{fin}\geq 1$ TeV. 

In Fig.~\ref{fig:events_sources}, we report the number of events for
the sources MGRO J2019+37, MGRO J1908+06, MGRO J2031+41 and for the
respective atmospheric background as a function of the measured
muon energy.  As described in the following
section, we will use these spectra for the calculation of the
confidence level at which a source could be excluded in the event of no
observation of any signal events, as well as the
statistical significance at which a source could be detected in the event
of a positive observation, depending on the specific values 
of $\alpha_\gamma$ and $E_{cut,\gamma}$.

\begin{table}[!t]
\centering
\begin{tabular}{|l || c | c | c | c |}
\hline
Source & 
R.A.~[hh mm ss] & Dec.~[dd mm ss] & $\sigma_{\rm ext}$ \\ [1ex]
\hline
MGRO J2019+37 & 20 18 35.03 & +36 50 00.0 & 0.75$^\circ$\\[6pt]
MGRO J1908+06 & 19 07 54 & +06 16 07 & 0.34$^\circ$\\[6pt]
MGRO J2031+41 & 20 29 38.4 & +41 11 24 & 1.10$^\circ$\\\hline
\end{tabular}
\caption{Position of the sources in right ascension and declination 
and extensions  of the sources~\cite{Abdo:2012jg,Aharonian:2009je}. }
\label{tab:sources_position}
\end{table}

\section{\label{sec:results} Results}

\subsection{\label{sec:sources} Milagro sources}
As mentioned before the first extraterrestrial neutrino flux
observed~\cite{Aartsen:2013jdh} by IceCube consists of 28 events 
with no event originating from nearby the Milagro sources.  
If
this continues to be the case in the following years of operation,
IceCube will be able to impose a bound on the probability of these
sources to be galactic PeVatrons. We refer to this probability as the
Confidence Level of exclusion of a given source. Conversely if this is
not the case and neutrinos are observed from the source direction in
the amount expected by the estimates presented in the previous
section IceCube will be able to establish the source as a galactic
PeVatron with some probability.  We refer to this as the Statistical
Significance of discovery.  We quantify these two statistical tests
for the three Milagro sources as a function of the detector exposure
and for a range of values of the source parameters $\alpha_\gamma$ and
$E_{cut,\gamma}$ as we describe next.

To estimate the Confidence Level of exclusion of a given source
we use as observable the total number of events with 
$E_\mu^{fin}\geq 1$ TeV without binning in energy. Generically,
for the small statistics expected, the use of the total number
of events yields a better rejection power than the use of  
the energy spectrum. We also studied the dependence of the 
Confidence Level of exclusion on the  minimum $E_\mu^{fin}$
and found that as long as  $E_\mu^{fin}$ was not so large that
the number of expected background events
became very small the  Confidence Level of exclusion
(C.L. from now on) did not change much. 

Following standard techniques 
\cite{Junk:1999kv,Read:2000ru,ATLAS:2011tau,Beringer:1900zz}, 
we define C.L. as 
\begin{equation}
C.L. = \frac{P_{(s+b)}}{1-P_b}\,.
\end{equation}
where $P_{(s+b)}$ and $P_b$ are the p-values for the signal plus
background and background only hypothesis of the data. 
Note that the denominator is present to avoid penalizing models to
which one has little or no sensitivity. If $C.L. \leq \alpha$, a
specific source is excluded with $(1- \alpha)$ confidence level.

To obtain  $P_{(s+b)}$ and $P_b$ we first  generate a large
sample of event numbers that are Poisson distributed around the total
expected number of background events and estimate what one could 
expect the  ``data'' of IceCube to be in the absence of any signal,
$\mathcal{N}_{D}$,  as their median.  
Knowing the theoretical expectations, for signal plus
background, $\mathcal{Y}_{(b+s)}$, and background only,
$\mathcal{Y}_{b}$, we  construct the data likelihood for signal plus
background, $\mathcal{L}_{(s+b),D}$, and background only,  
$\mathcal{L}_{b,D}$. 

We then generate a large number of experimental results
$N_{exp}$, that are Poisson distributed around the expected total number of
signal plus background (background only) events for which we  
can compute the corresponding likelihoods of signal+background  
$\mathcal{L}_{(s+b),J}$ (background only,  $\mathcal{L}_{b,J}$)
with $J=1 \dots N_{exp}$ and count  the number of results,
$N_{pos}(s+b)$ ($N_{pos}(b)$)  for which 
$-2 \ln \mathcal{L}_{(s+b),J} > -2 \ln \mathcal{L}_{(s+b),D}$
($-2 \ln \mathcal{L}_{b,J} > -2 \ln \mathcal{L}_{b,D})$
so 
\begin{eqnarray}
P_{s+b}=\frac{N_{pos}(s+b)}{N_{exp}}\,,& \;\;\;\;\;\;
& 
1-P_{b}= \frac{N_{pos}(b)}{N_{exp}}\,.
\end{eqnarray}

\begin{figure}
\centering
\begin{tabular}{lcr}
\includegraphics[width=0.32\textwidth]{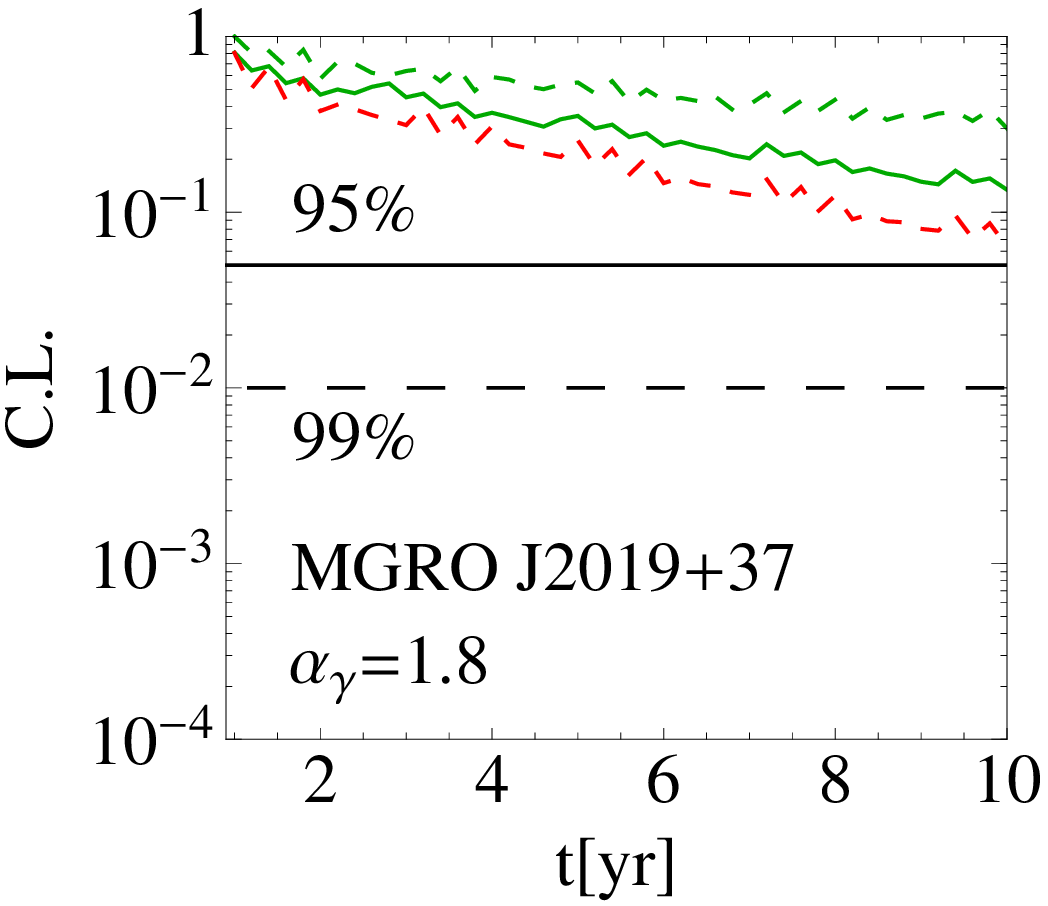} &
\includegraphics[width=0.32\textwidth]{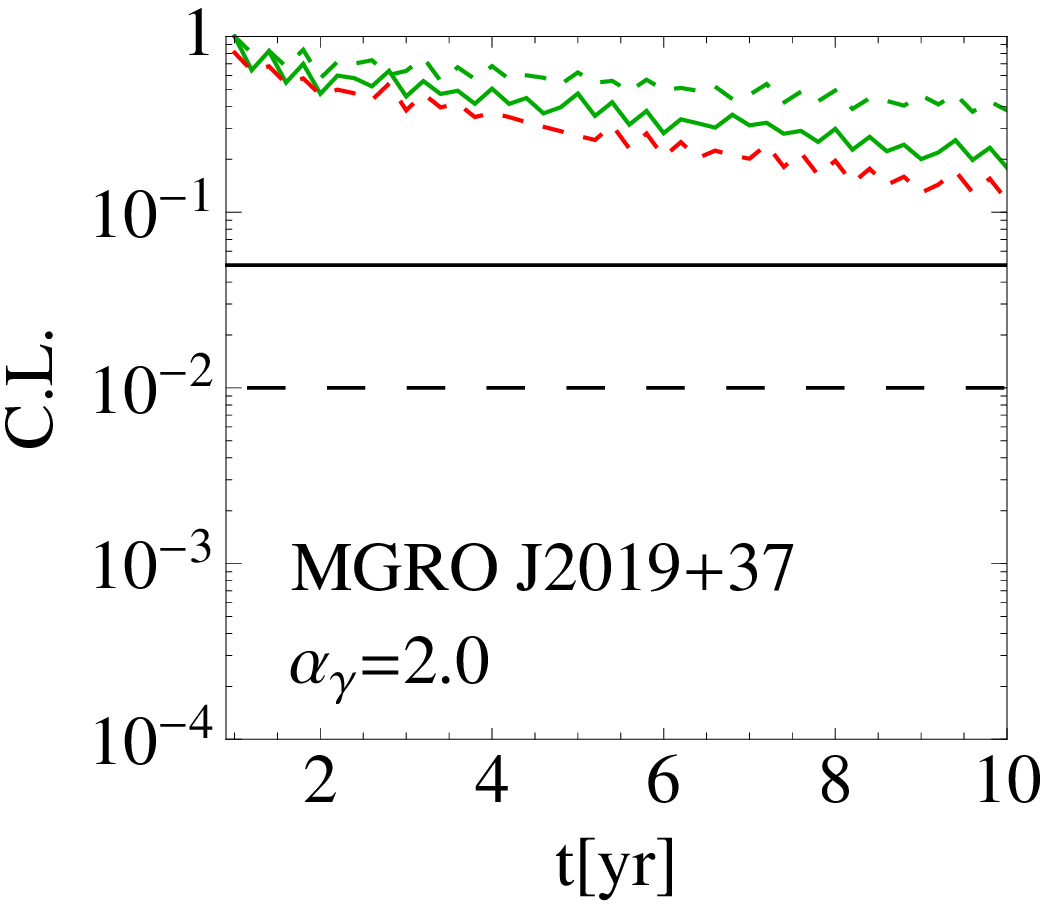} &
\includegraphics[width=0.32\textwidth]{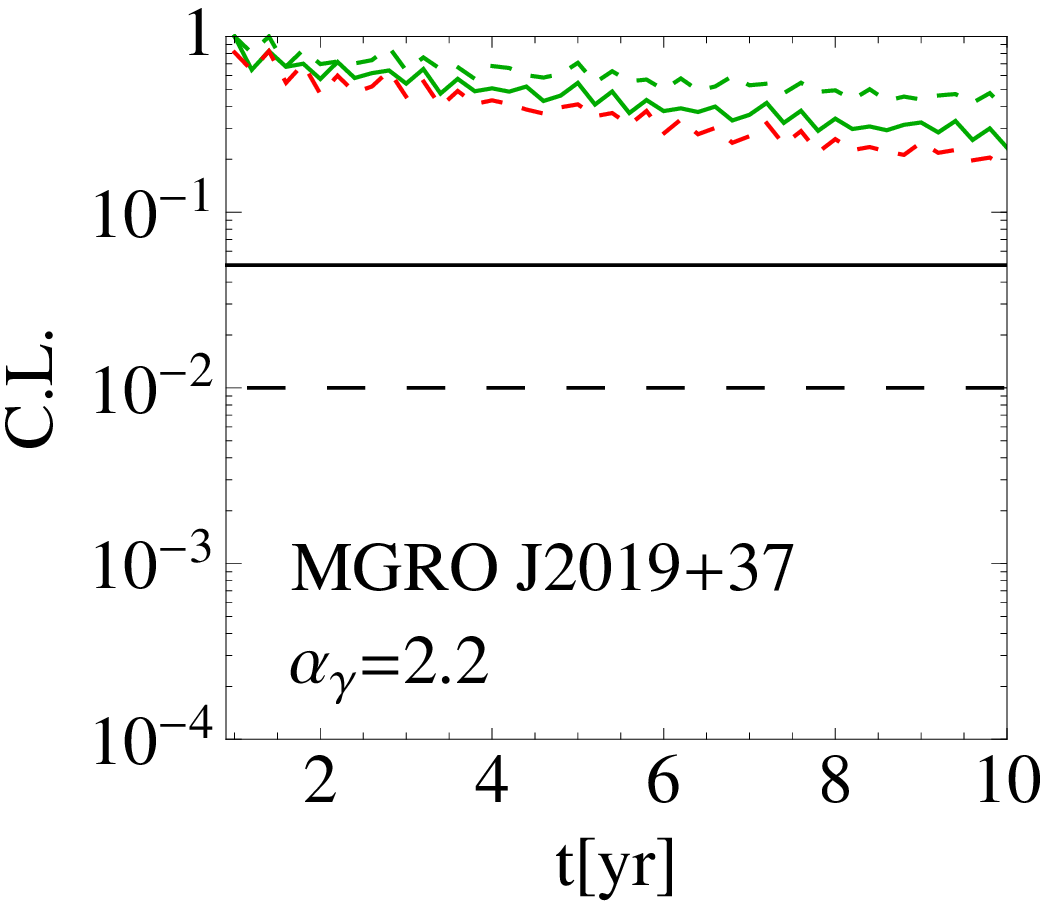} \\
\includegraphics[width=0.32\textwidth]{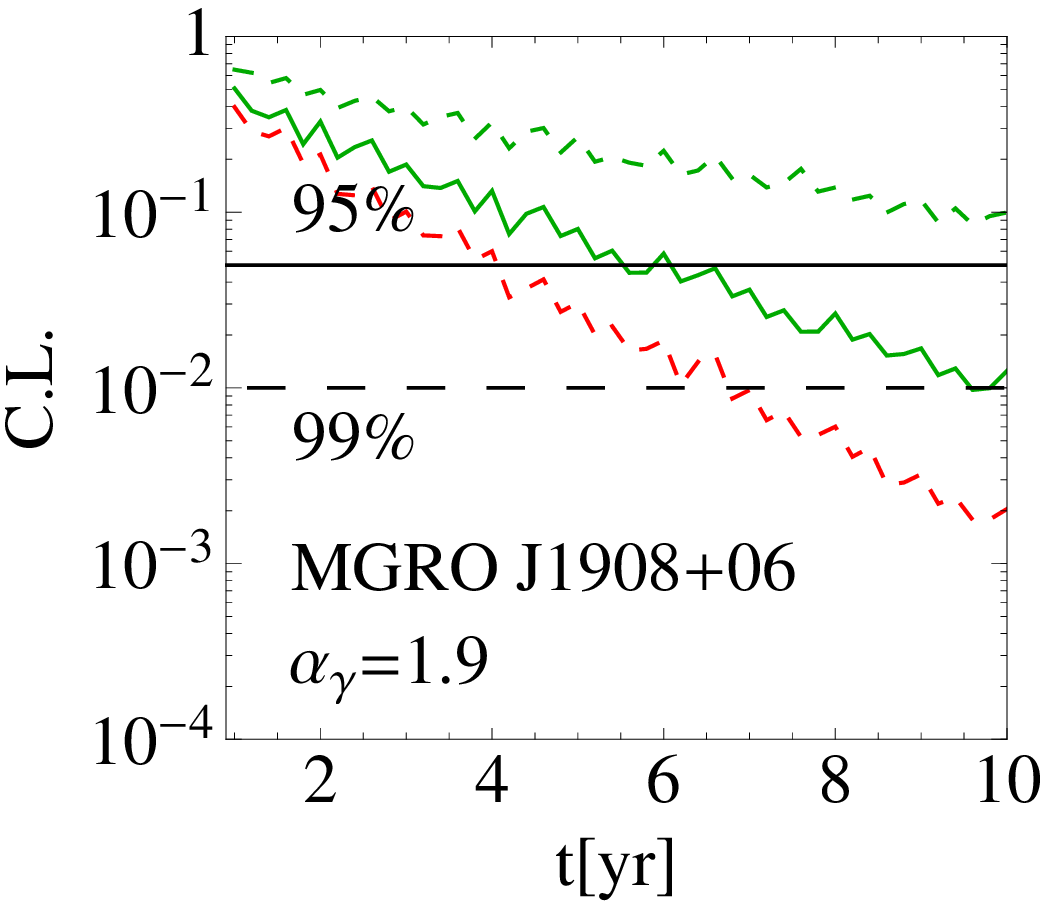} &
\includegraphics[width=0.32\textwidth]{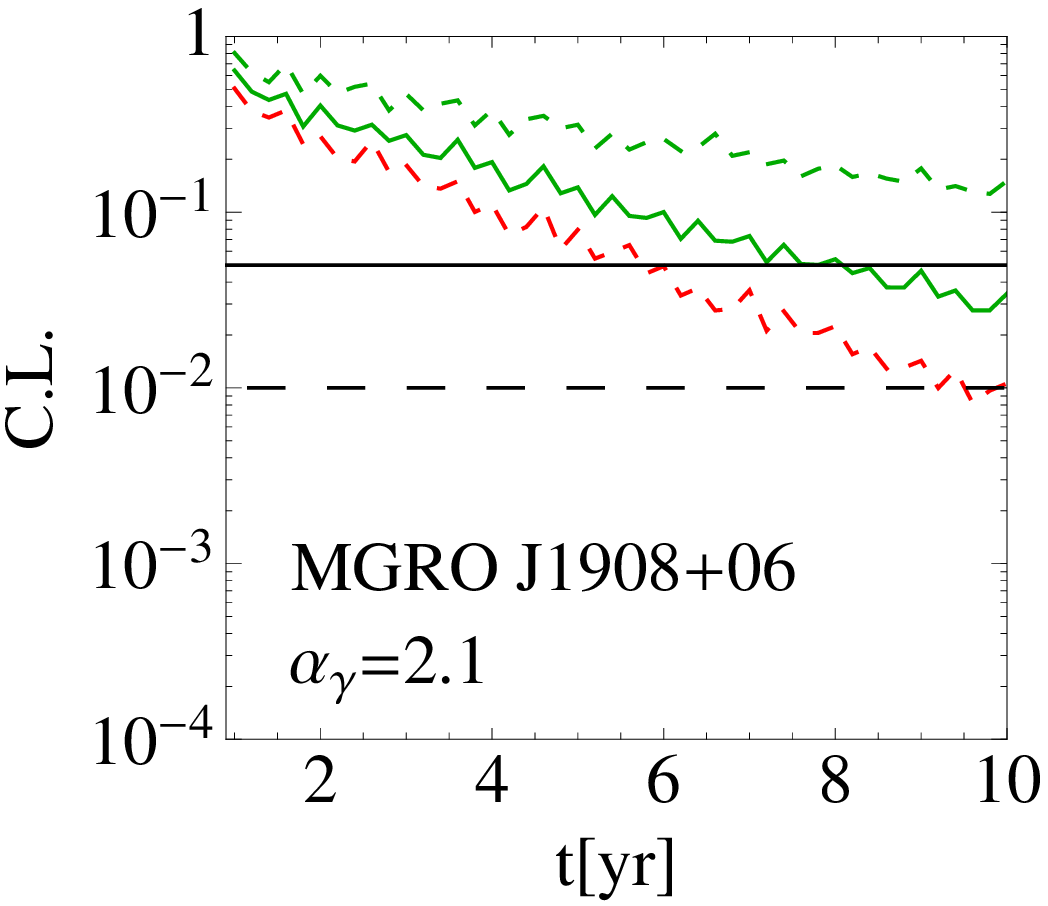} &
\includegraphics[width=0.32\textwidth]{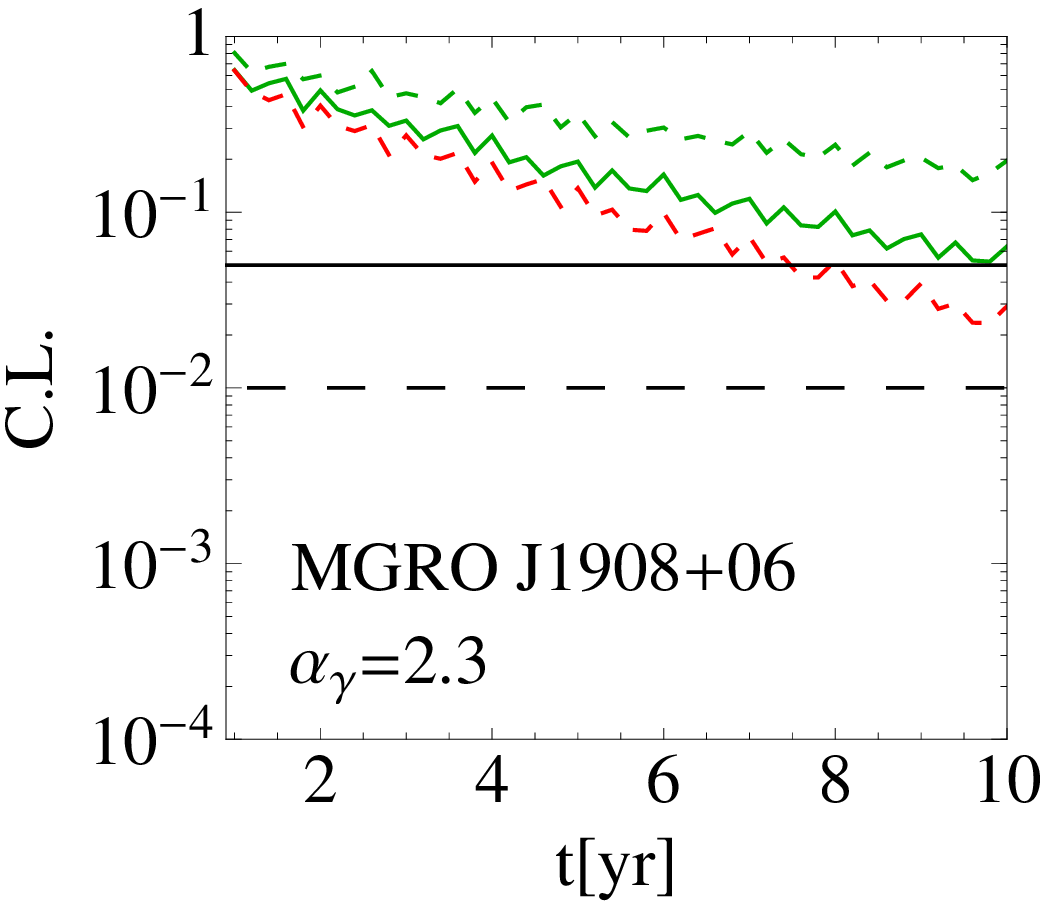} \\
\includegraphics[width=0.32\textwidth]{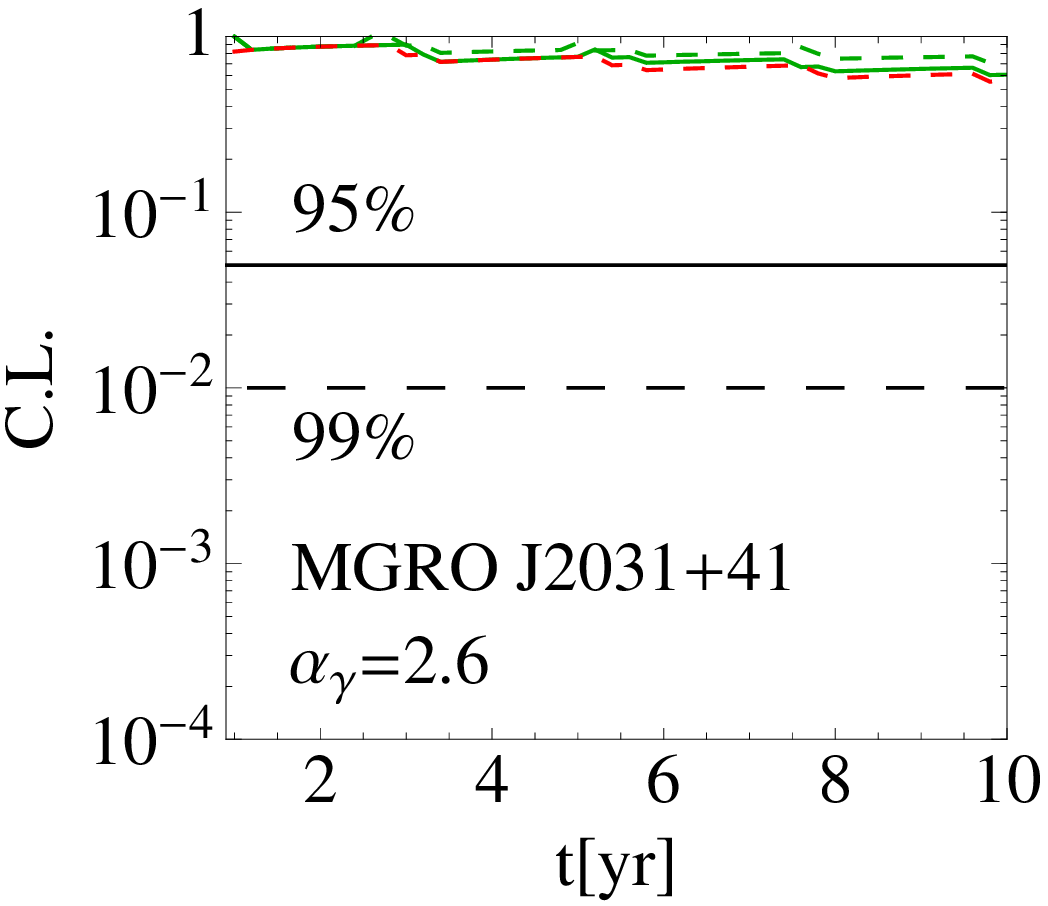} &
\includegraphics[width=0.32\textwidth]{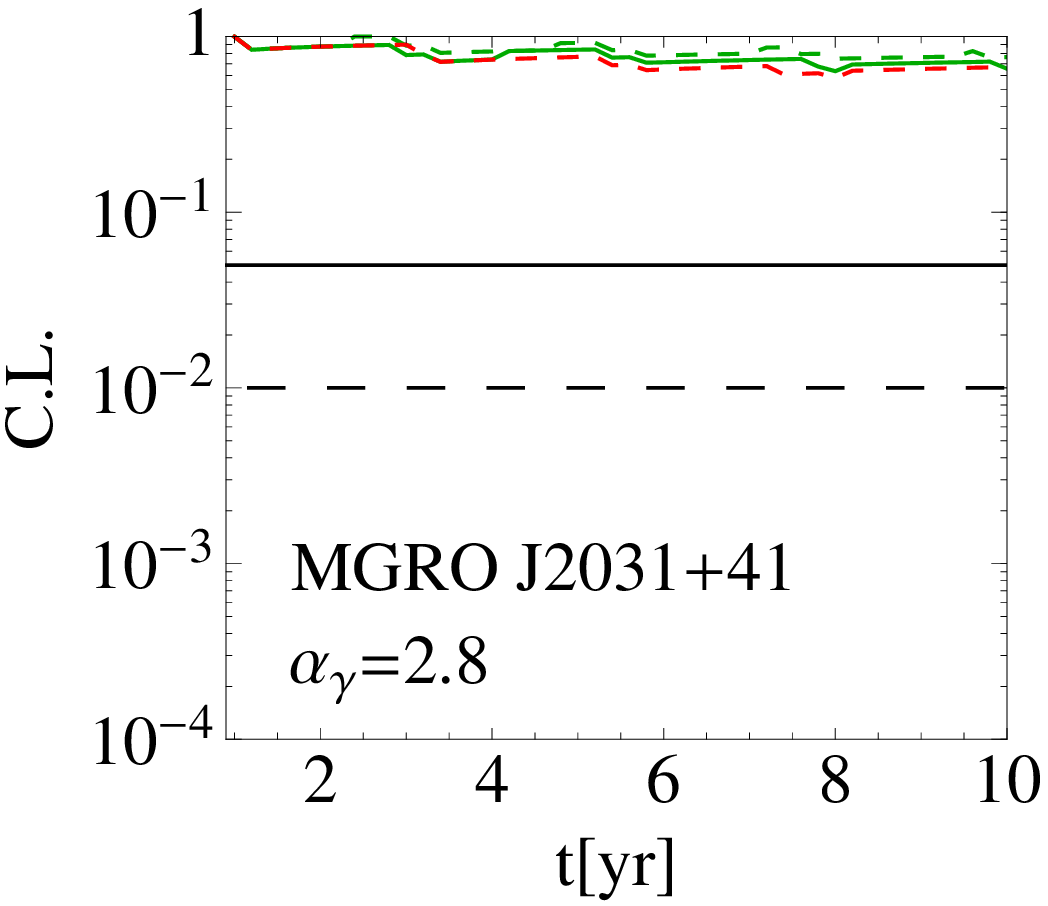} &
\includegraphics[width=0.32\textwidth]{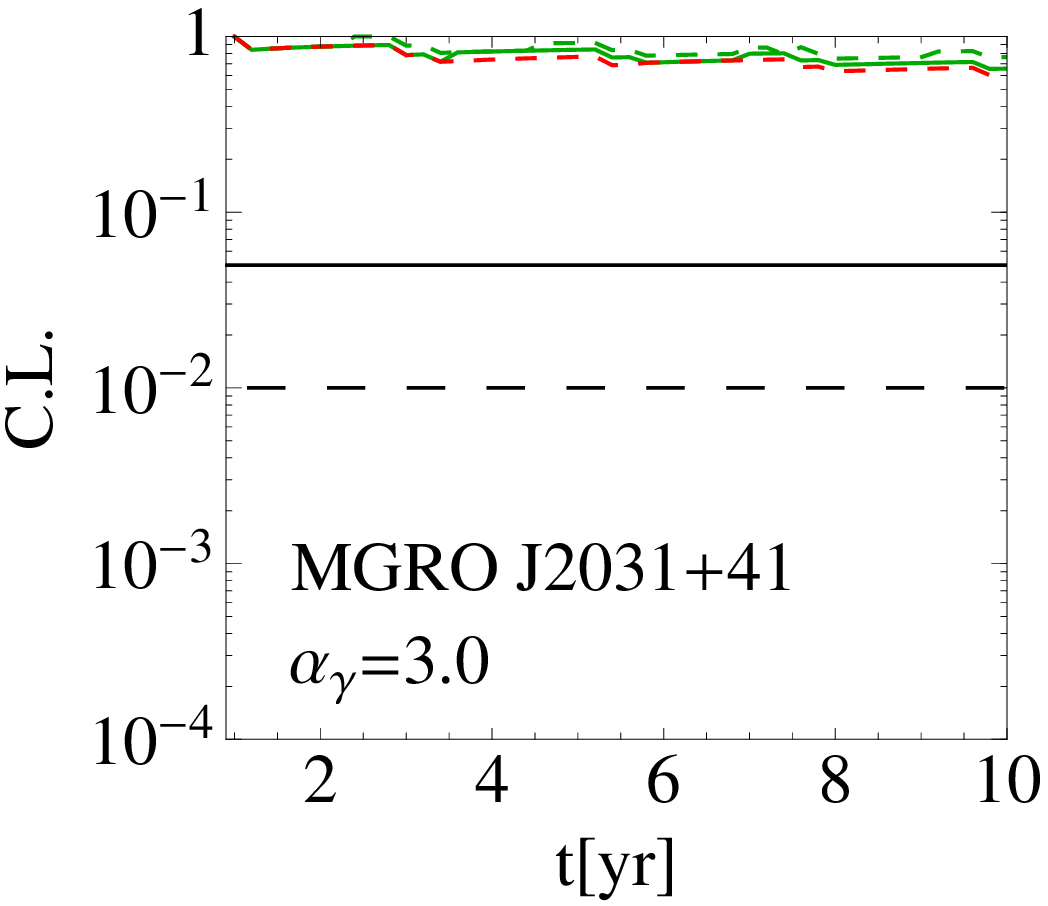}
\end{tabular}
\caption{\label{fig:sum_CL_sources} 
Confidence level at which a source could be excluded as a function of
time. For each sources, we have considered different values of
$\alpha_\gamma$. The green
dashed, solid green and red dashed lines refer to the values of $E_{cut,\gamma}$ 
as reported in Table~\ref{tab:sources_norm}, from smallest to biggest values. The observed spectra
are obtained as the median of random generated values around the
background.}
\end{figure}

The results for the expected C.L. are presented in
Fig.~\ref{fig:sum_CL_sources} for the three sources considered in the
paper and for the different values of $\alpha_\gamma$ and
$E_{cut,\gamma}$, as reported in Table~\ref{tab:sources_norm}.
We find that the parameters of the sources MGRO J2019+37 
and MGRO J2031+41 will be difficul to constrain at 95\%~C.L. in less than 10 years. 
Instead, for MGRO J1908+06,
in roughly 4 (7) years the values $\alpha_\gamma=1.9$ and
$E_{cut,\gamma}=45$~TeV could be excluded at 95\% (99\%) confidence level.
Considering $\alpha_\gamma=2.1$ and $E_{cut,\gamma}=45$~TeV, an
exclusion at 95\% (99\%) confidence level is possible in 6 (10) years, while for 
$\alpha_\gamma=2.3$ and $E_{cut,\gamma}=45$~TeV, roughly 8 years are necessary
for an exclusion at 95\% confidence level. For a lower value of 
$E_{cut, \gamma}$ of 30~TeV, an exclusion at 95\% (99\%) confidence level 
is possible in 7 (10) years for $\alpha_\gamma=1.9$, in 8 years for $\alpha_\gamma=2.1$ 
and in 10 years for $\alpha_\gamma=2.3$.

\begin{figure}
\centering
\begin{tabular}{lcr}
\includegraphics[width=0.32\textwidth]{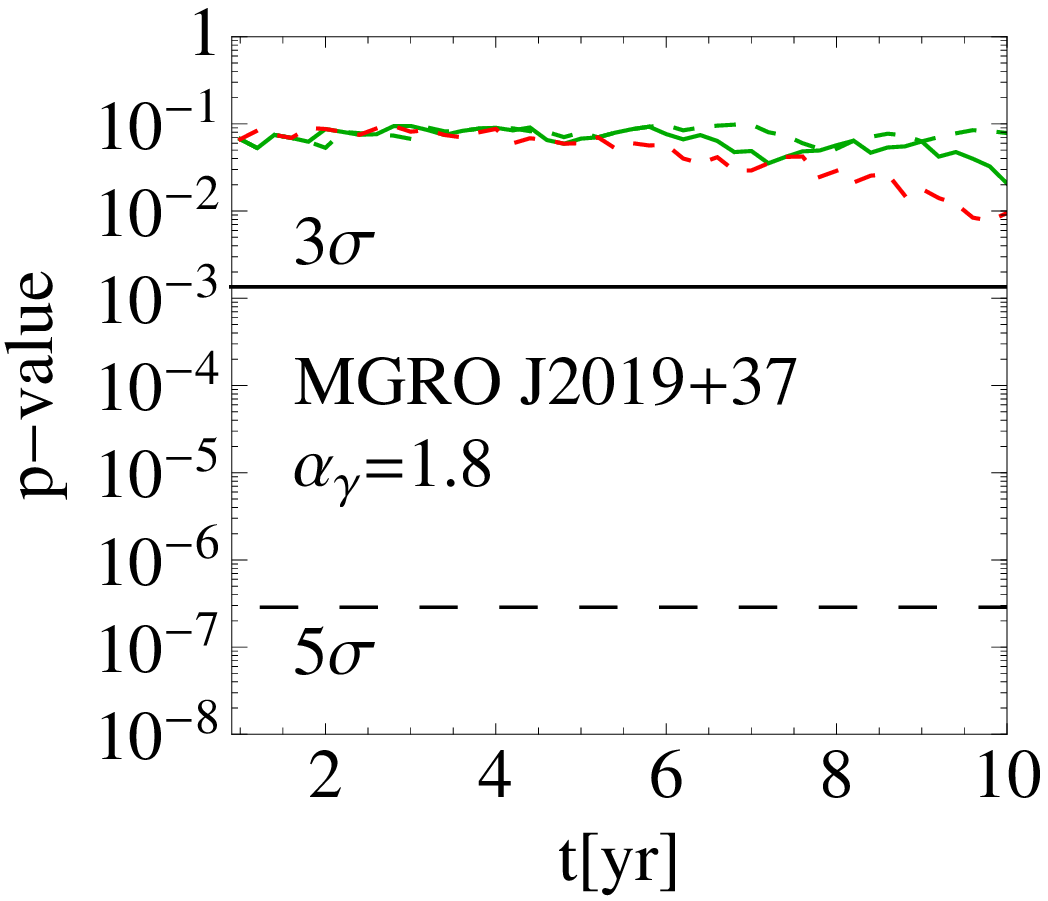} &
\includegraphics[width=0.32\textwidth]{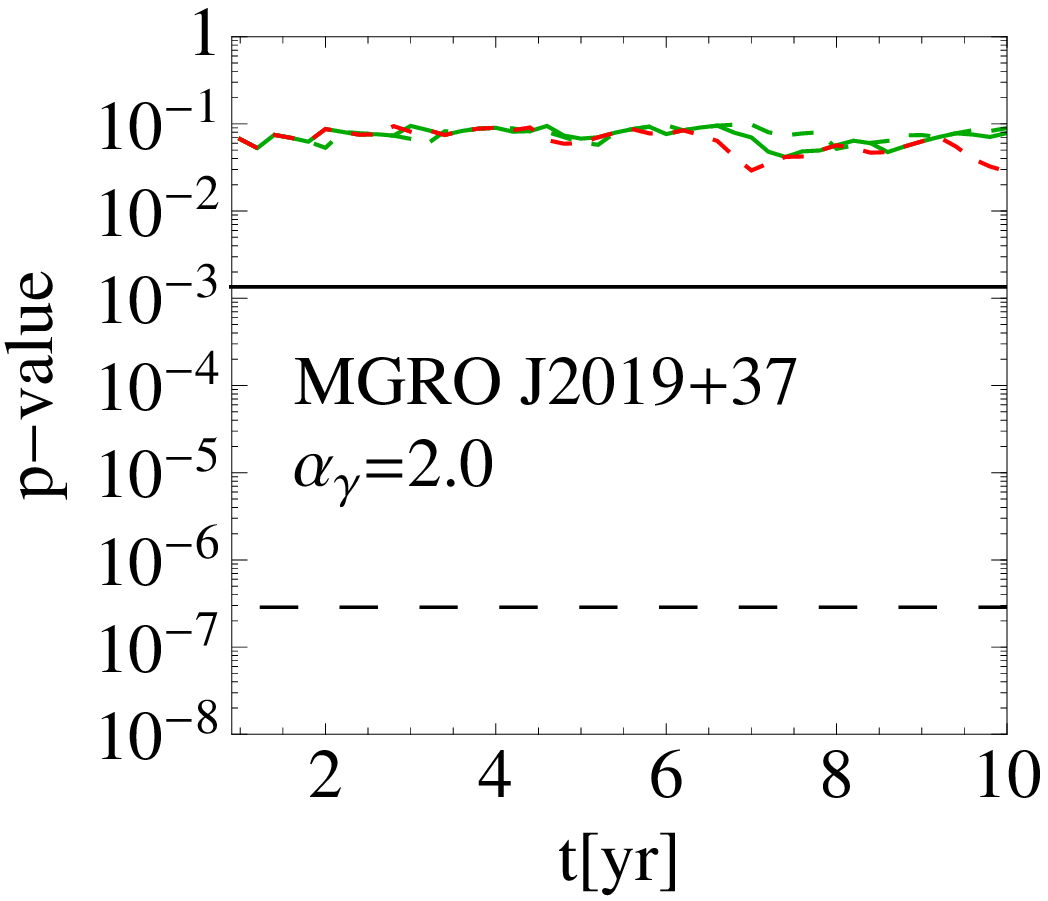} &
\includegraphics[width=0.32\textwidth]{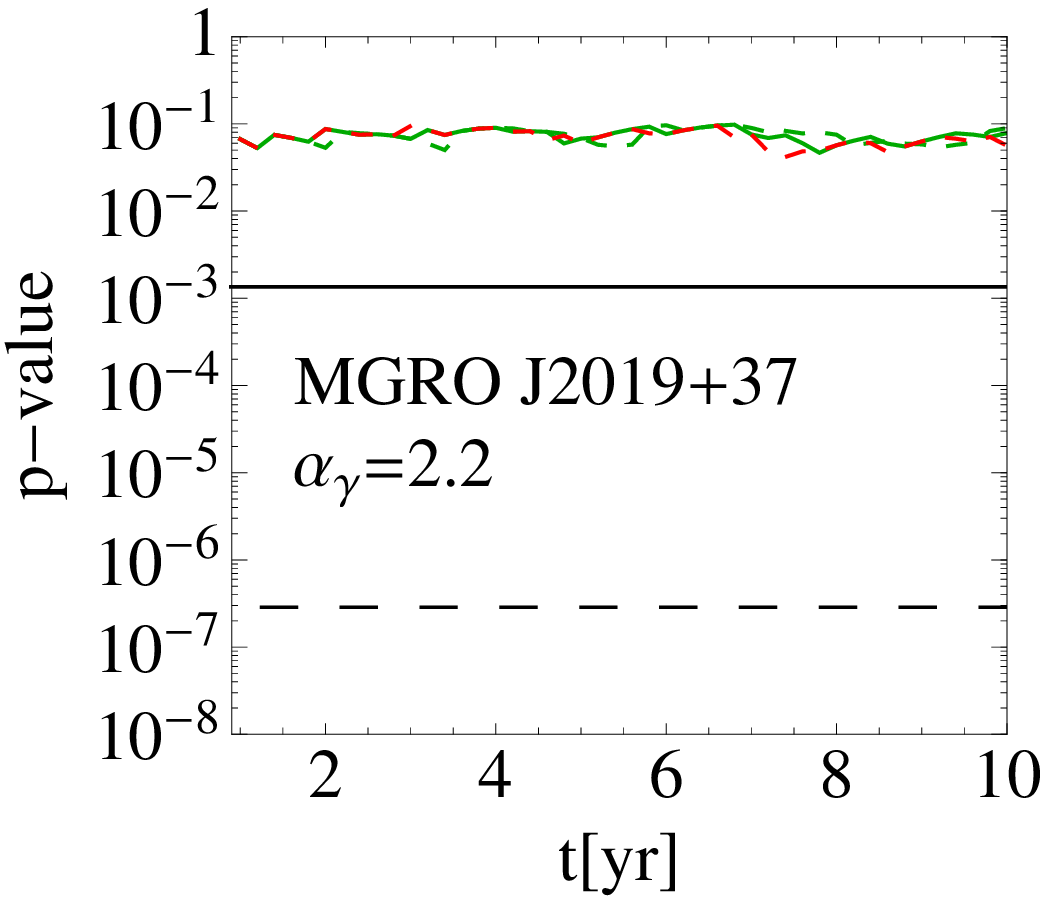} \\
\includegraphics[width=0.32\textwidth]{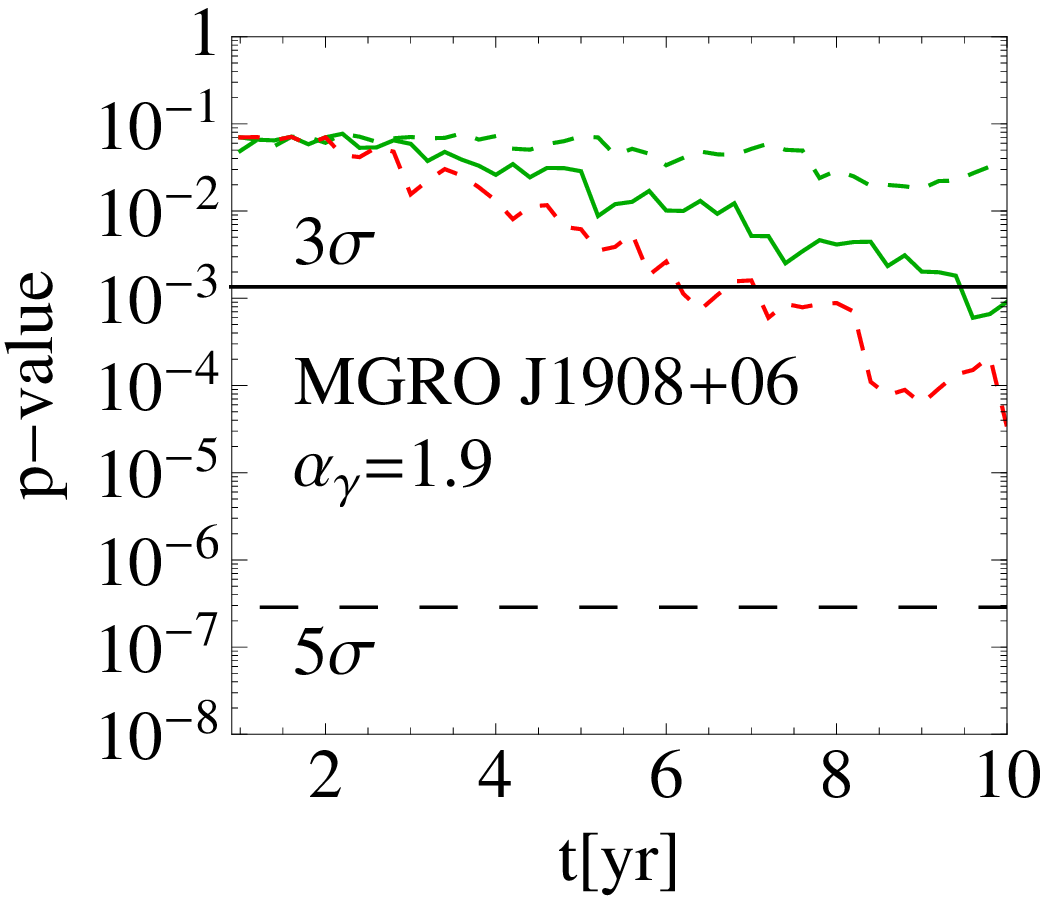} &
\includegraphics[width=0.32\textwidth]{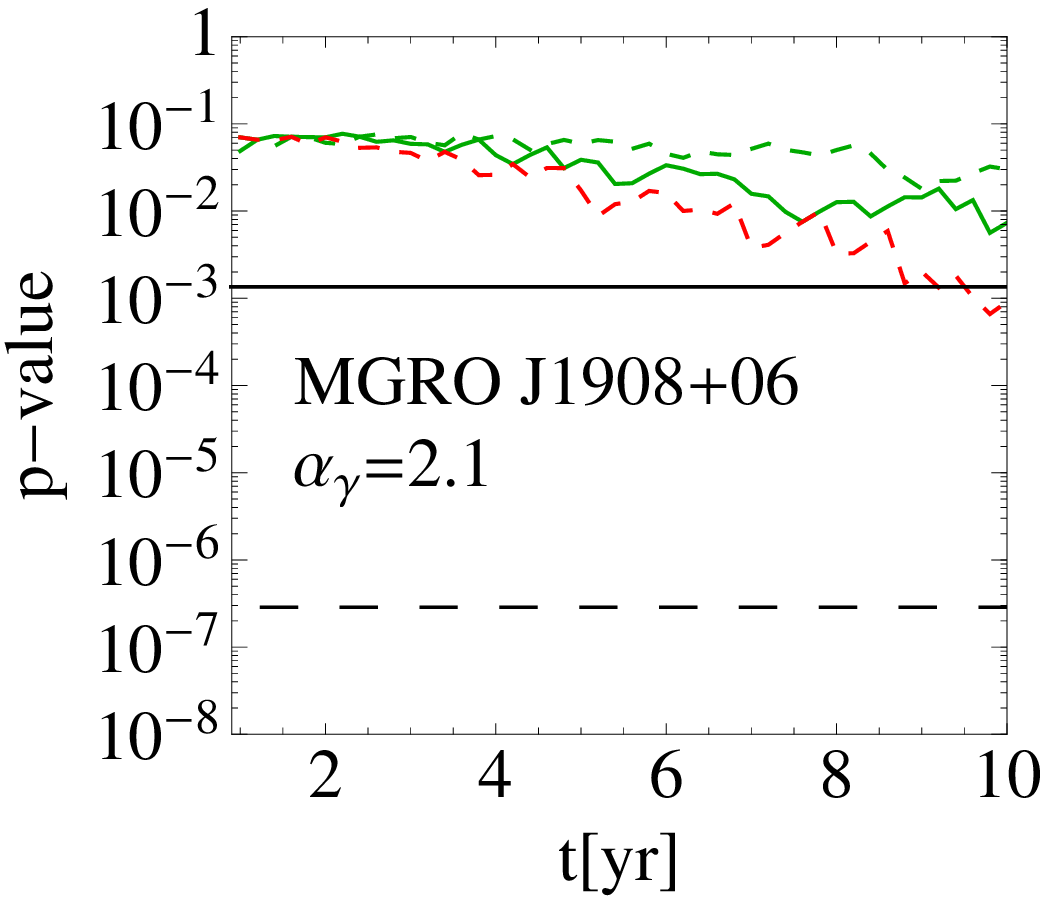} &
\includegraphics[width=0.32\textwidth]{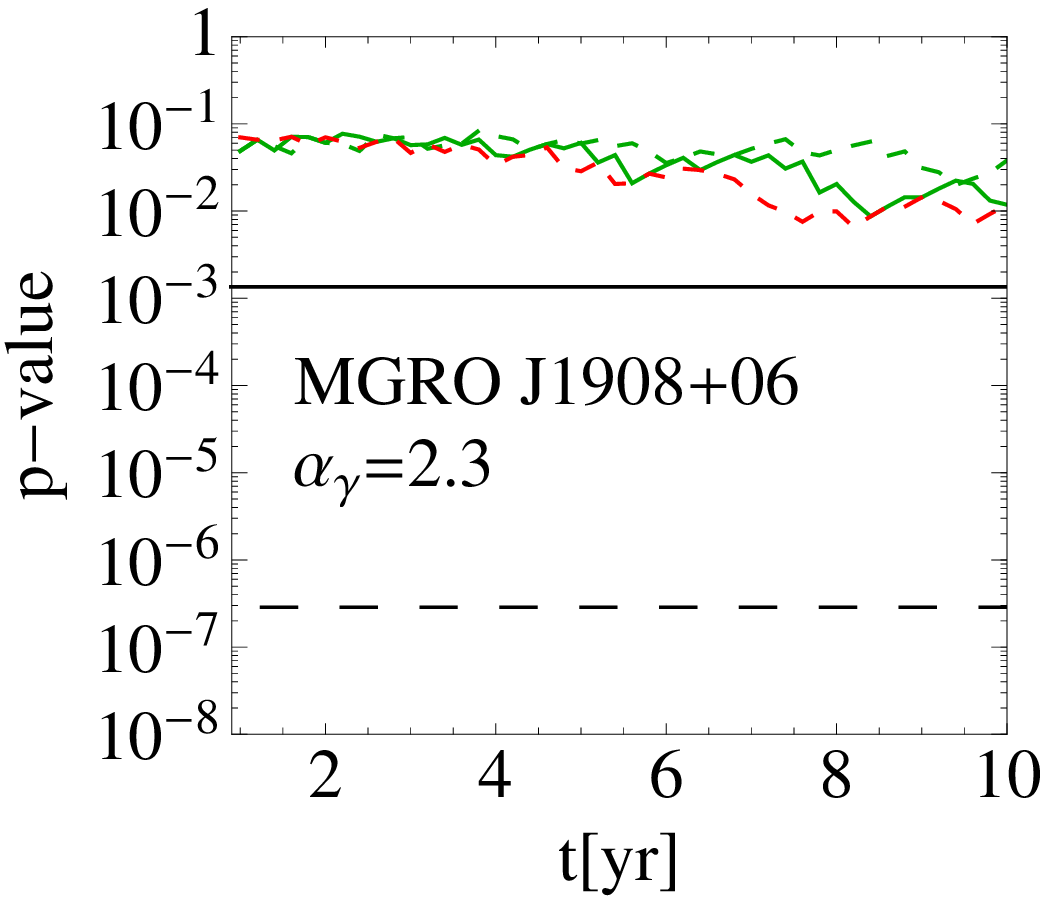} \\
\includegraphics[width=0.32\textwidth]{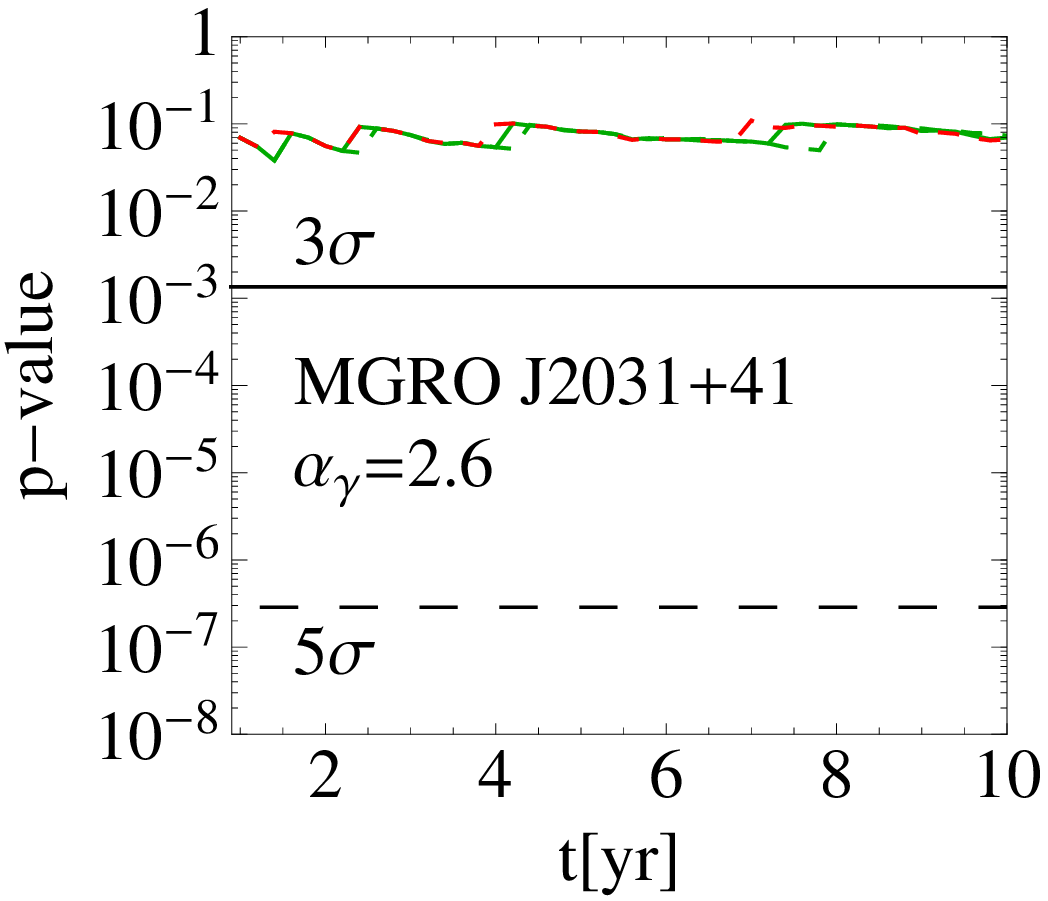} &
\includegraphics[width=0.32\textwidth]{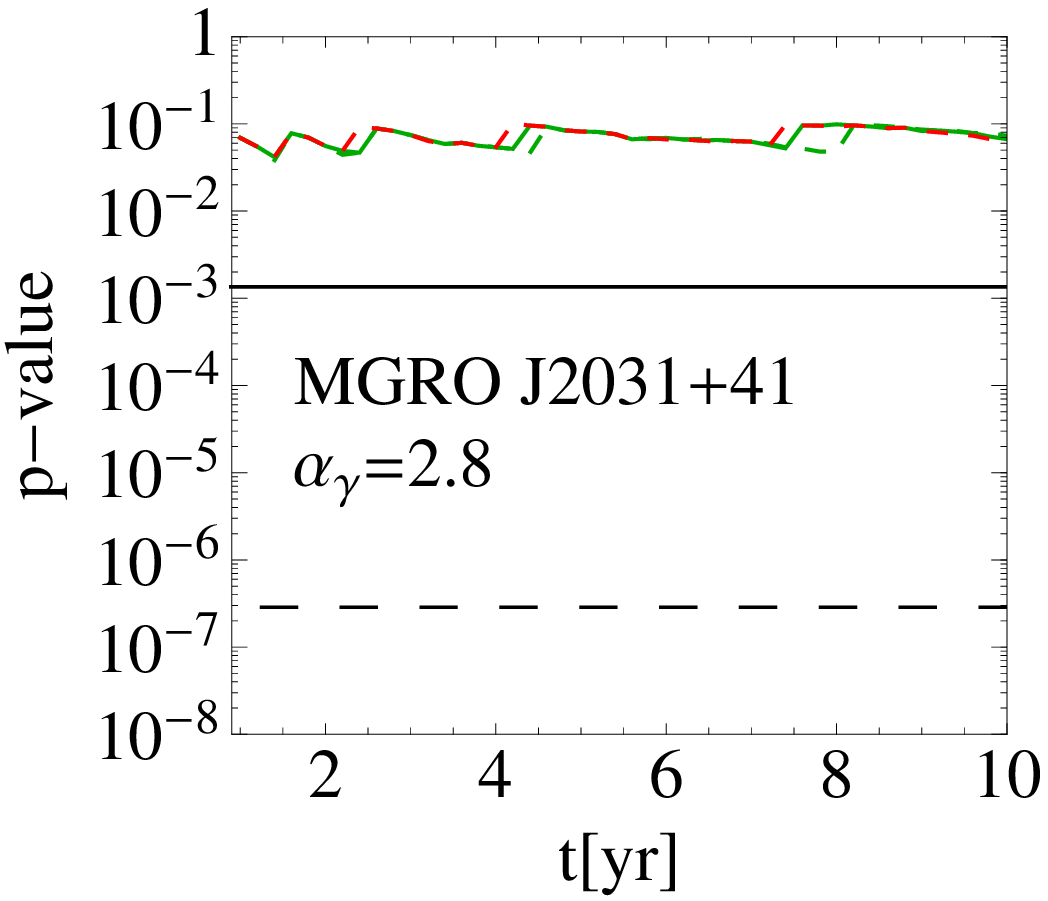} &
\includegraphics[width=0.32\textwidth]{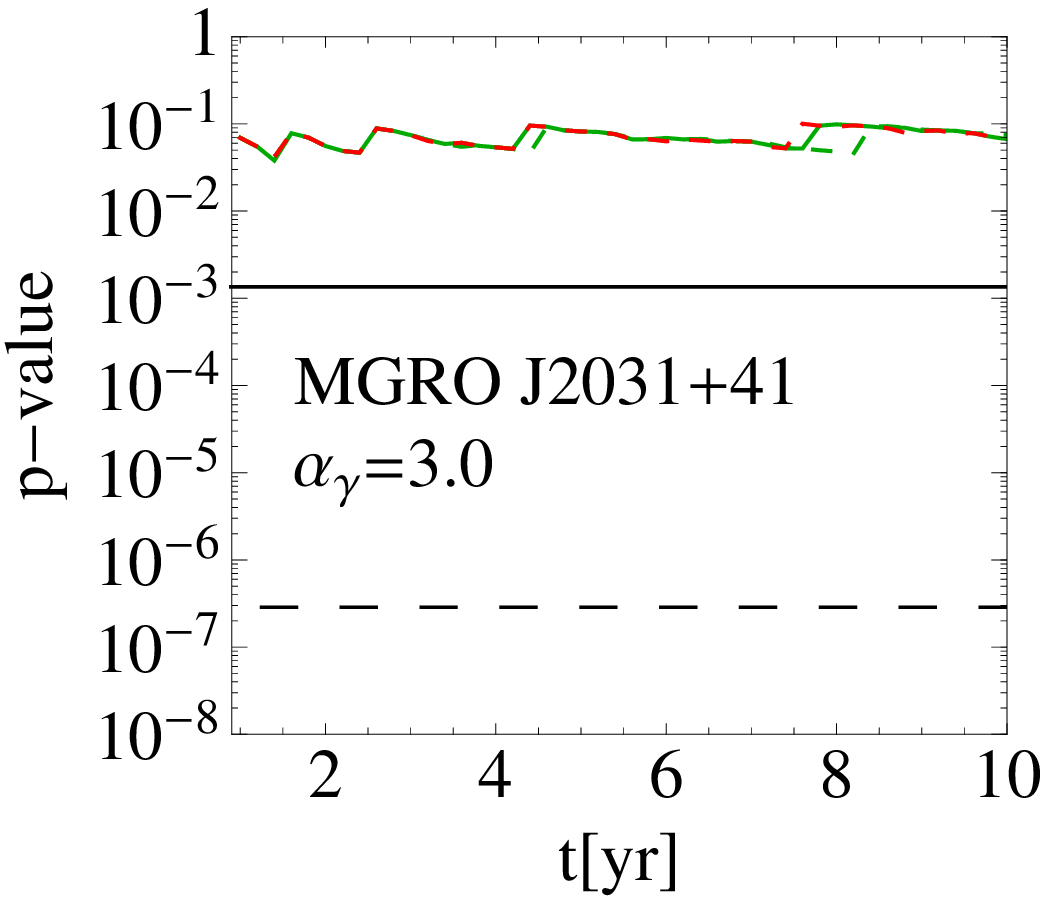} 
\end{tabular}
\caption{\label{fig:SS} Statistical significance at which a source
could be detected as a function of time. For each source, we
considered different values of $\alpha_\gamma$. The green
dashed, solid green and red dashed lines refer to the values of $E_{cut,\gamma}$ 
as reported in Table~\ref{tab:sources_norm}, from smallest to biggest values. 
The observed spectra are obtained as the median of
random generated values around the signal plus background. }
\end{figure}

Conversely when estimating the Statistical Significance of discovery 
we find higher  sensitivity to the signal when using the full spectral 
information of expected signal and background events. 
We compute the statistical significance of observing a signal 
as the background-only p-value, considering the analytic
expression reported in Ref.~\cite{ATLAS:2011tau}: 
\begin{equation}
p_{\rm value}=\frac{1}{2}\left[ 1-{\rm{erf}} 
\left( \sqrt{q_0^{obs}/2} \right) \right]\,,
\end{equation}
where $q_0^{obs}$ is defined as 
\begin{equation}
q_0^{obs} \equiv -2 \ln \mathcal{L}_{b,D}= 2 \sum_i \left( Y_{b,i} - N_{D,i} + 
N_{D,i} \ln \left( \frac{N_{D,i}}{Y_{b,i}}\right)\right)\,,
\end{equation}
with $i$ running over the different energy bins. In this case  
$N_{D,i}$ is the estimated experimental data  in bin $i$
--generated as the median of a large sample of event numbers 
that are Poisson distributed around the expectation of signal plus
background--, $Y_{b,i}$ is the theoretical expectation for the
background hypothesis. 

The results for the statistical significance are reported in 
Fig.~\ref{fig:SS}. From the figure we read that the sources MGRO J2019+37 
and MGRO J2031+41 will be difficul to detect at 3$\sigma$ level in less than 10 years, 
considering the parameters $\alpha_\gamma$ and $E_{cut,\gamma}$, as reported 
in Table~\ref{tab:sources_norm}. 
The source MGRO J1908+06, instead, could be detected at 3$\sigma$ in roughly 7 year for 
$\alpha_\gamma=1.9$ and $E_{cut,\gamma}=45$~TeV and in 9 years for 
$\alpha_\gamma=2.1$. 
Recently, the authors of Ref.~\cite{Tchernin:2013wfa} found that the IceCube detector would be able to 
detect the sources MGRO J2019+37 and MGRO J2031+41 of the Cygnus region only after 20 years of exposure. 
This is consistent with the results of our study. 
Moreover, a recent analysis done by the IceCube collaboration, with 3 years data, has revealed no detection 
for the sources MGRO J2019+37 and MGRO J1908+06~\cite{Aartsen:2013uuv}. This is consistent with our findings, since 
in 3 years, it is not possible to discovery these sources at 3$\sigma$ level, see Fig.~\ref{fig:SS}.

\begin{table}[!t]
\centering
\begin{tabular}{|c || c | c | c | c |}
\hline
 Source & \#~of~yrs for C.L.~@~95\% & \#~of~yrs for p-value~@~3$\sigma$\\ [1ex]
\hline
MGRO J2019+37 & $\alpha_\gamma= \{1.8,~2.0,~2.2 \}~\rightarrow$  & $\alpha_\gamma= \{1.8,~2.0,~2.2 \}~\rightarrow$ \\
 & \#~of~yrs:  $\{>10,~>10,~>10\}$ &  \#~of~yrs: $\{>10,~>10,~>10\}$ \\[6pt]
MGRO J1908+06 & $\alpha_\gamma= \{1.9,~2.1,~2.3 \}~\rightarrow$  & $\alpha_\gamma= \{1.9,~2.1,~2.3 \}~\rightarrow$ \\
 & \#~of~yrs: $\{4,~6,~8\}$ & \#~of~yrs: $\{7,~9,~>10\}$ \\[6pt]
MGRO J2031+41 & $\alpha_\gamma= \{2.6,~2.8,~3.0 \}~\rightarrow$  & $\alpha_\gamma= \{2.6,~2.8,~3.0 \}~\rightarrow$ \\
 & \#~of~yrs: $\{>10,~>10,~>10\}$ & \#~of~yrs: $\{>10,~>10,~>10\}$ \\\hline
\end{tabular}
\caption{Results on the C.L. and p-value for the three different sources that 
could be obtained in {\it{less than ten years}}. We have considered $E_{cut,\gamma}$=45~TeV for 
the first and second source and $E_{cut,\gamma}$=300~TeV for the third source.}
\label{tab:results}
\end{table}

\subsection{IceCube evidence for cosmic neutrinos: 
are (some of) the neutrinos of Galactic origin?}
If cosmic accelerators are the origin of the extraterrestrial flux of
neutrinos recently observed by IceCube~\cite{Aartsen:2013jdh}, then the
neutrinos have most likely been produced in proton-photon or
proton-proton interactions with radiation or gas, either at the
acceleration site or along the path traveled by cosmic rays to
Earth. The fraction of energy transferred to pions is about 20\%
(50\%) for $p\gamma$ ($pp$), respectively, and each of the three
neutrinos from the decay chain $\pi^+\to\mu^+\nu_\mu$ and $\mu^+\to
e^+\nu_e\bar\nu_\mu$ carries on average one quarter of the pion
energy. Hence, the cosmic rays producing the excess neutrinos have
energies of tens of PeV, well above the knee in the spectrum. It is
tantalizingly close to the energy of
$100$~PeV~\cite{Apel:2013dga,IceCube:2013wda} where the spectrum
displays a rich structure, sometimes referred to as the ``iron knee."
While these cosmic rays are commonly categorized as Galactic, with the
transition to the extragalactic population at the ankle in the
spectrum at $3 \sim 4$\,EeV, one cannot rule out a subdominant
contribution of PeV neutrinos of extragalactic origin. IceCube
neutrinos may give us information on the much-debated transition
energy.

\begin{figure}
\centering
\begin{tabular}{lr}
\includegraphics[width=0.4\textwidth]{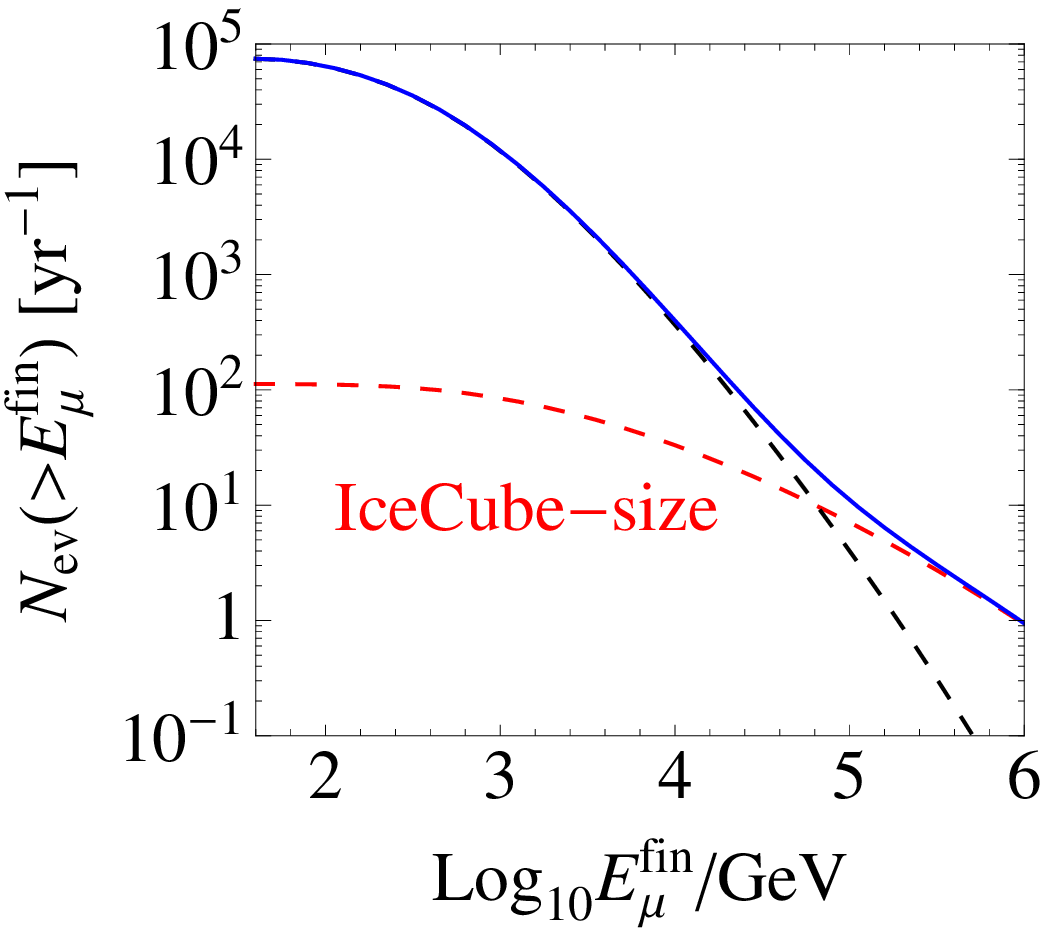}&
\includegraphics[width=0.4\textwidth]{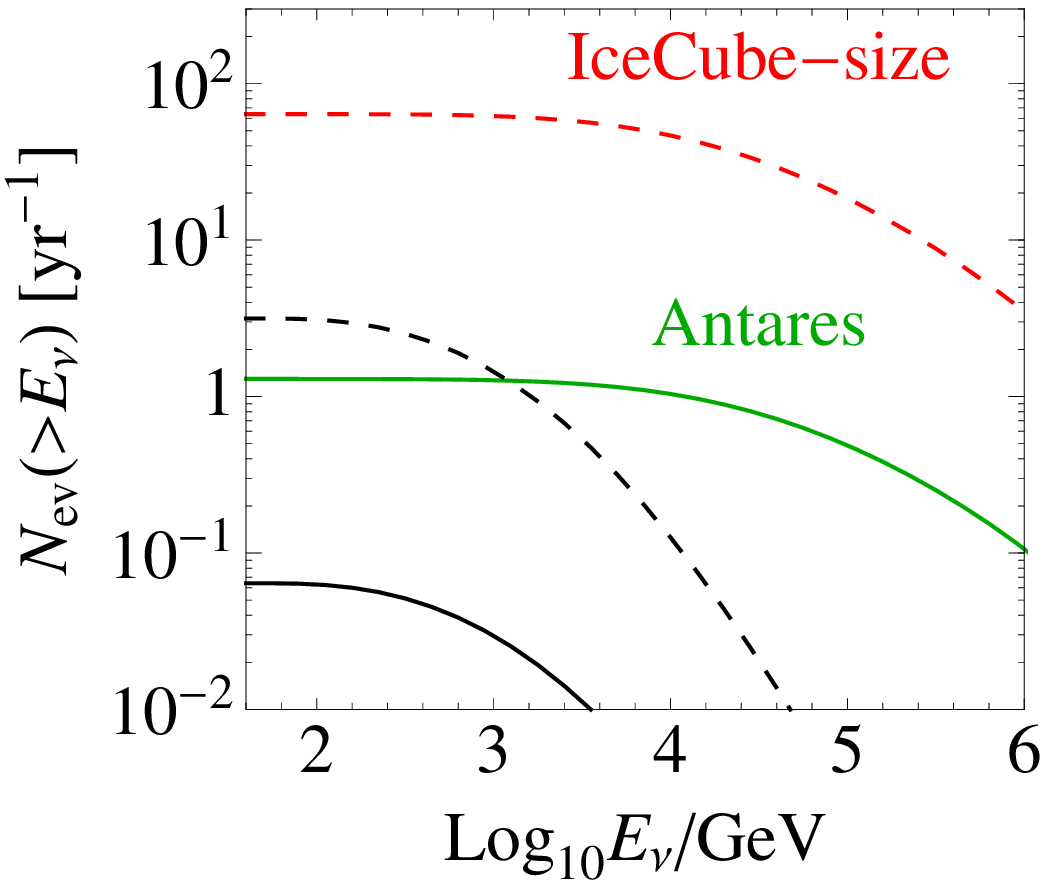}
\end{tabular}
\caption{\label{fig:GC} 
Left panel: Events for the diffuse flux of Eq.(\ref{eq:dif1fflux}) 
as a function of the muon energy. We report the number of events 
for an IceCube-size detector in the north hemisphere
(red-dashed line) together with
the respective atmospheric background (black-dashed). The sum of the 
events from the diffuse flux and from the atmospheric background is reported 
with a blue line. 
Right panel: Events from Galactic Center source as a
function of the neutrino energy.  We report the number of events and
the background for an IceCube-size detector in the north hemisphere
(red-dashed line) and for Antares (green-solid line), together with
the respective atmospheric background: black-dashed for IceCube and
black-solid for Antares. For the atmospheric background we have
integrated over a steradian $\Omega=\pi (1.6 \sigma)^2$, with $\sigma
= 0.4^\circ$. Note that the number of events for the Galactic Center 
source and the background events are corrected for the fact that the center of the 
Galaxy is only visible 68\% of the time for a Mediterranean detector. } 
\end{figure}

The acceptance of the starting event analysis producing the first
evidence for an extraterrestrial neutrino flux is such that the signal
consist mostly of electron and tau neutrinos originating in the
Southern hemisphere. While their energy can be reconstructed to 15\%,
their direction is only measured to $10\sim15$ degrees. In contrast, a
detector in the Mediterranean views the Southern hemisphere through
the Earth and therefore has sensitivity to muon neutrinos that can be
reconstructed with sub-degree precision. For illustration, an IceCube
detector cloned and positioned in the Mediterranean would observe a
diffuse flux of 71 muon-neutrinos\footnote{This is likely to be an overestimate because the effective area for a 
diffuse analysis, which typically requires stronger cuts on the data, is smaller 
than the point source area used here. In the case of IceCube this correction is 
close to a factor of two.} per year with energy in excess of
45\,TeV for muon neutrino flux of:
\begin{equation}
E^2_\nu \frac{dN_{\nu_\mu + \bar{\nu}_\mu}}{dE_\nu} = 
1.2 \times 10^{-11} \quad {\rm TeV}~{\rm cm}^{-2}~{\rm s}^{-1}~{\rm sr}^{-1}\,.
\label{eq:dif1fflux}
\end{equation}
We here assumed a 1:1:1 distribution of flavors which is consistent
with observation. Above 45\,TeV, observed event rates should be 
dominated by the flux of the source, providing a sky map with little 
background. In the left panel of Fig.~\ref{fig:GC}, we report the 
number of muon events as a function of the muon energy for the diffuse 
flux of Eq.~\eqref{eq:dif1fflux} and the respective atmospheric background. 

As mentioned in the introduction a cluster of seven events is observed 
close to the center of the Galaxy. If these events are  
originated from a point source, the corresponding
flux would be
\begin{equation}
E^2_\nu \frac{dN_{\nu_\mu + \bar{\nu}_\mu}}{dE_\nu} = 6 \times 10^{-11}\quad {\rm TeV}~{\rm cm}^{-2}~{\rm s}^{-1}\,,
\label{eq:GCpoint}
\end{equation}
corresponding to roughly 41 events per year above 45\,TeV in neutrino energy. 
This number is not corrected for the fact that the center of the Galaxy is only 
visible 68\% of the time for a Mediterranean detector. We introduced this correction in 
the corresponding figure, see the right panel of Fig.~\ref{fig:GC}. Note that the flux is simply estimated by 
multiplying the diffuse flux in Eq.(\ref{eq:dif1fflux}) 
by $4\pi \times 7/17.4$ where we correct for the fact that 7 events 
out of 17.4 (we subtract the 10.6 of background estimation to the measured 28 events) are clustered 
around the Galactic Center and that both samples consist mainly of
shower events. Note that the point source flux of Eq.(\ref{eq:GCpoint}) 
is compatible with the all-flavor flux reported in Eq.(2) of Ref.~\cite{Razzaque:2013uoa}, 
calculated for a 0.06 sr solid angle surrounding the Galactic Center. 

Both the diffuse and point source signals would be statistically
significant within one year. The operating Antares detector is a
factor of 40 smaller than the IceCube detector, and therefore the
IceCube excess only produces signals at the one-event level per
year. Larger event samples, especially of well-reconstructed muon
neutrinos, are likely to be the key to a conclusive identification of
the origin of the IceCube extraterrestrial flux. If the flux observed
in IceCube turns out to be isotropic, IceCube itself will observe the
same muon neutrino signals from the Northern hemisphere.

If, on the other hand, any of the IceCube events do originate from a Galactic
point source, IceCube itself should be able to observe the
accompanying PeV gamma rays.  The distance to the center of the Galaxy
corresponds to a single interaction length for a PeV photon
propagating in the microwave background. PeV photons are detected as
muon-poor showers triggered by IceTop leaving no muons in IceCube. The
level of point-source flux per neutrino flavor corresponding to one
out of the 28 events is given by
\begin{eqnarray}
E_\nu^2 \frac{dN_\nu}{dE_\nu}= 4\pi \,\frac{1}{28} \,1.2\times10^{-11}\,\rm cm^{-2}\,s^{-1}\,TeV  \nonumber \\
\simeq 5.4\times10^{-12}\,\rm cm^{-2}\,s^{-1}\,TeV\,,
\label{eq:psflux}
\end{eqnarray}
with the corresponding pionic photon flux a factor of 2 larger,
assuming $pp$ interactions; see above. This is a flux of
$\sim10^{-17}\,\rm cm^{-2}\,s^{-1}\,TeV^{-1}$ at 1\,PeV, well within
the gamma-ray sensitivity of the completed IceCube detector; see
Fig.~15 in \cite{Aartsen:2012gka}. In fact, the highest fluctuation in
a gamma-ray map obtained with one year of data collected with the
detector when it was half complete is in the direction of one of the
PeV neutrino events~\cite{Ahlers:2013xia}.

\section{Summary}
\label{sec:conclu} 
Recently, the IceCube detector reported evidence for extraterrestrial
neutrinos, at very high-energies.  Among the data collected between
May 2010 and May 2012, 28 neutrino events were detected with energies
between 20 and 1200 TeV, while from a background estimation roughly
10.6 events were expected. None of these events seems to 
originate from the nearby star-forming region in Cygnus, posing the
question of whether the observed gamma-ray sources in this region 
are indeed the postulated PeVatrons originating the galactic cosmic rays.

In view of these results and taking into account recent TeV gamma ray
measurements of the spectra of some of the sources, we have 
updated  the calculation of the neutrino expected from 
Milagro sources MGRO J2019+37, MGRO J1908+06 and MGRO J2031+41 
3 of the 6 sources used in the IceCube stacking analysis based on
references~\cite{GonzalezGarcia:2009jc,Halzen:2008zj,Kappes:2009zza}.
We have then estimated the confidence level with which IceCube
could rule out these sources as galactic PeVatrons, should the 
results continue being negative in the following years of operation,
or the statistical significance with which it can discover neutrinos  
from these sources if they are indeed galactic PeVatrons. 

In particular, we find that the parameters of the sources MGRO J2019+37 
and MGRO J2031+41 will be difficul to constrain at 95\%~C.L. in less than 10 years. 
For MGRO J1908+06, instead, in roughly 4 years the values $\alpha_\gamma=1.9$ and
$E_{cut,\gamma}=45$~TeV could be excluded at 95\% confidence level.
Increasing the value of $\alpha_\gamma$ to 2.1, an exclusion is possible in 6 years, while for 
$\alpha_\gamma=2.3$ roughly 8 years are necessary. 
Considering the statistic significance, instead, we found that the sources MGRO J2019+37 
and MGRO J2031+41 will be difficul to detect at 3$\sigma$ level in less than 10 years, 
considering the parameters $\alpha_\gamma$ and $E_{cut,\gamma}$ of our analysis. 
The source MGRO J1908+06 could be detected at 3$\sigma$ in roughly 7 year for 
$\alpha_\gamma=1.9$ and $E_{cut,\gamma}=45$~TeV and in 9 years for 
$\alpha_\gamma=2.1$.

Moreover, among the 28 events, a hot spot of 7 shower events is
evident at a position close to the Galactic Center. We studied the
possibility of detecting the point-source flux associated to this hot
spot by a kilometer-scale detector in the Northern hemisphere and by
the Antares detector.


\section*{Acknowledgments}
This work is supported by USA-NSF grant PHY-09-6739, by CUR
Generalitat de Catalunya grant 2009SGR502 by MICINN FPA2010-20807 and
consolider-ingenio 2010 program grants CUP (CSD-2008-00037) and CPAN,
and by EU grant FP7 ITN INVISIBLES (Marie Curie Actions
PITN-GA-2011-289442). F.H. is supported in part by the U.S. National
Science Foundation under Grants No.~OPP-0236449 and PHY-0969061and by
the University of Wisconsin Research Committee with funds granted by
the Wisconsin Alumni Research Foundation.


\bibliographystyle{elsarticle-num}
\bibliography{pevatrons}

\begin{thebibliography}{10}
\expandafter\ifx\csname url\endcsname\relax
  \def\url#1{\texttt{#1}}\fi
\expandafter\ifx\csname urlprefix\endcsname\relax\def\urlprefix{URL }\fi
\expandafter\ifx\csname href\endcsname\relax
  \def\href#1#2{#2} \def\path#1{#1}\fi

\bibitem{BaadeAndZwicky}
W.~Baade, F.~Zwicky, On super-novae, Proceedings of the National Academy of
  Science 20 (1934) 254--259.

\bibitem{Ackermann:2013wqa}
M.~Ackermann, et~al., {Detection of the Characteristic Pion-Decay Signature in
  Supernova Remnants}, Science 339 (2013) 807.
\newblock \href {http://arxiv.org/abs/1302.3307} {\path{arXiv:1302.3307}}.

\bibitem{DjannatiAtai:2007ny}
A.~Djannati-Atai, E.~Ona-Wilhelmi, M.~Renaud, S.~Hoppe, {H.E.S.S. Galactic
  Plane Survey unveils a Milagro Hotspot. }\href
  {http://arxiv.org/abs/0710.2418} {\path{arXiv:0710.2418}}.

\bibitem{Albert:2008yk}
J.~Albert, et~al., {MAGIC observations of the unidentified TeV gamma-ray source
  TeV J2032+4130}, Astrophys.J. 675 (2008) L25--L28.
\newblock \href {http://arxiv.org/abs/0801.2391} {\path{arXiv:0801.2391}}.

\bibitem{Gabici:2007qb}
S.~Gabici, F.~A. Aharonian, {Searching for galactic cosmic ray pevatrons with
  multi-TeV gamma rays and neutrinos. }\href {http://arxiv.org/abs/0705.3011}
  {\path{arXiv:0705.3011}}.

\bibitem{Halzen:2008zj}
F.~Halzen, A.~Kappes, A.~O'Murchadha, {Prospects for identifying the sources of
  the Galactic cosmic rays with IceCube}, Phys.Rev. D78 (2008) 063004.
\newblock \href {http://arxiv.org/abs/0803.0314} {\path{arXiv:0803.0314}}.

\bibitem{GonzalezGarcia:2009jc}
M.~Gonzalez-Garcia, F.~Halzen, S.~Mohapatra, {Identifying Galactic PeVatrons
  with Neutrinos}, Astropart.Phys. 31 (2009) 437--444.
\newblock \href {http://arxiv.org/abs/0902.1176} {\path{arXiv:0902.1176}}.

\bibitem{Vissani:2011ea}
F.~Vissani, F.~Aharonian, {Galactic Sources of High-Energy Neutrinos:
  Highlights}, Nucl.Instrum.Meth. A692 (2012) 5--12.
\newblock \href {http://arxiv.org/abs/1112.3911} {\path{arXiv:1112.3911}}.

\bibitem{Vissani:2011vg}
F.~Vissani, F.~Aharonian, N.~Sahakyan, {On the Detectability of High-Energy
  Galactic Neutrino Sources}, Astropart.Phys. 34 (2011) 778--783.
\newblock \href {http://arxiv.org/abs/1101.4842} {\path{arXiv:1101.4842}}.

\bibitem{Aartsen:2013jdh}
M.~Aartsen, et~al., {Evidence for High-Energy Extraterrestrial Neutrinos at the
  IceCube Detector}, Science 342~(6161) (2013) 1242856.
\newblock \href {http://arxiv.org/abs/1311.5238} {\path{arXiv:1311.5238}}.

\bibitem{ARGO-YBJ:2012goa}
B.~Bartoli, et~al., {Observation of the TeV gamma-ray source MGRO J1908+06 with
  ARGO-YBJ. }\href {http://arxiv.org/abs/1207.6280} {\path{arXiv:1207.6280}}.

\bibitem{Abdo:2012jg}
A.~Abdo, U.~Abeysekara, B.~Allen, T.~Aune, D.~Berley, et~al., {Spectrum and
  Morphology of the Two Brightest Milagro Sources in the Cygnus Region: MGRO
  J2019+37 and MGRO J2031+41}, Astrophys.J. 753 (2012) 159.
\newblock \href {http://arxiv.org/abs/1202.0846} {\path{arXiv:1202.0846}}.

\bibitem{Kappes:2009zza}
A.~Kappes, F.~Halzen, A.~O.~Murchadha, {Prospects of identifying the sources of
  the galactic cosmic rays with IceCube}, Nucl.Instrum.Meth. A602 (2009)
  117--119.

\bibitem{Fox:2013oza}
D.~Fox, K.~Kashiyama, P.~Mészarós, {Sub-PeV Neutrinos from TeV Unidentified
  Sources in the Galaxy}, Astrophys.J. 774 (2013) 74.
\newblock \href {http://arxiv.org/abs/1305.6606} {\path{arXiv:1305.6606}}.

\bibitem{Kistler:2006hp}
M.~D. Kistler, J.~F. Beacom, {Guaranteed and Prospective Galactic TeV Neutrino
  Sources}, Phys.Rev. D74 (2006) 063007.
\newblock \href {http://arxiv.org/abs/astro-ph/0607082}
  {\path{arXiv:astro-ph/0607082}}.

\bibitem{Kelner:2006tc}
S.~Kelner, F.~A. Aharonian, V.~Bugayov, {Energy spectra of gamma-rays,
  electrons and neutrinos produced at proton-proton interactions in the very
  high energy regime}, Phys.Rev. D74 (2006) 034018.
\newblock \href {http://arxiv.org/abs/astro-ph/0606058}
  {\path{arXiv:astro-ph/0606058}}.

\bibitem{Kappes:2006fg}
A.~Kappes, J.~Hinton, C.~Stegmann, F.~A. Aharonian, {Potential Neutrino Signals
  from Galactic Gamma-Ray Sources}, Astrophys.J. 656 (2007) 870--896.
\newblock \href {http://arxiv.org/abs/astro-ph/0607286}
  {\path{arXiv:astro-ph/0607286}}.

\bibitem{Abdo:2007ad}
A.~Abdo, B.~T. Allen, D.~Berley, S.~Casanova, C.~Chen, et~al., {TeV Gamma-Ray
  Sources from a Survey of the Galactic Plane with Milagro}, Astrophys.J. 664
  (2007) L91--L94.
\newblock \href {http://arxiv.org/abs/0705.0707} {\path{arXiv:0705.0707}}.

\bibitem{Abdo:2009ku}
A.~Abdo, B.~Allen, T.~Aune, D.~Berley, C.~Chen, et~al., {Milagro Observations
  of TeV Emission from Galactic Sources in the Fermi Bright Source List},
  Astrophys.J. 700 (2009) L127--L131.
\newblock \href {http://arxiv.org/abs/0904.1018} {\path{arXiv:0904.1018}}.

\bibitem{Bartoli:2012tj}
B.~Bartoli, P.~Bernardini, X.~Bi, C.~Bleve, I.~Bolognino, et~al., {Observation
  of TeV gamma rays from the Cygnus region with the ARGO-YBJ experiment},
  Astrophys.J. 745 (2012) L22.
\newblock \href {http://arxiv.org/abs/1201.1973} {\path{arXiv:1201.1973}}.

\bibitem{Beacom:2007yu}
J.~F. Beacom, M.~D. Kistler, {Dissecting the Cygnus Region with TeV Gamma Rays
  and Neutrinos}, Phys.Rev. D75 (2007) 083001.
\newblock \href {http://arxiv.org/abs/astro-ph/0701751}
  {\path{arXiv:astro-ph/0701751}}.

\bibitem{Smith:2010yn}
A.~J. Smith, {A Survey of Fermi Catalog Sources using Data from the Milagro
  Gamma-Ray Observatory. }\href {http://arxiv.org/abs/1001.3695}
  {\path{arXiv:1001.3695}}.

\bibitem{Aharonian:2009je}
F.~Aharonian, {Detection of Very High Energy radiation from HESS J1908+063
  confirms the Milagro unidentified source MGRO J1908+06. }\href
  {http://arxiv.org/abs/0904.3409} {\path{arXiv:0904.3409}}.

\bibitem{Aharonian:2005ex}
F.~Aharonian, et~al., {The Unidentified TeV source (TeV J2032+4130) and
  surrounding field: Final HEGRA IACT-system results}, Astron.Astrophys. 431
  (2005) 197--202.
\newblock \href {http://arxiv.org/abs/astro-ph/0501667}
  {\path{arXiv:astro-ph/0501667}}.

\bibitem{Lang:2004bk}
M.~J. Lang, D.~Carter-Lewis, D.~Fegan, S.~Fegan, A.~Hillas, et~al., {Evidence
  for TeV gamma ray emission from TeV j2032+4130 in whipple archival data},
  Astron.Astrophys. 423 (2004) 415--419.
\newblock \href {http://arxiv.org/abs/astro-ph/0405513}
  {\path{arXiv:astro-ph/0405513}}.

\bibitem{Abdo:2010ht}
A.~Abdo, A.~Abdo, {PSR J1907+0602: A Radio-Faint Gamma-Ray Pulsar Powering a
  Bright TeV Pulsar Wind Nebula}, Astrophys.J. 711 (2010) 64--74.
\newblock \href {http://arxiv.org/abs/1001.0792} {\path{arXiv:1001.0792}}.

\bibitem{Tchernin:2013wfa}
C.~Tchernin, J.~Aguilar, A.~Neronov, T.~Montaruli, {Neutrino signal from
  extended Galactic sources in IceCube}\href {http://arxiv.org/abs/1305.4113}
  {\path{arXiv:1305.4113}}.

\bibitem{GonzalezGarcia:2005xw}
M.~Gonzalez-Garcia, F.~Halzen, M.~Maltoni, {Physics reach of high-energy and
  high-statistics icecube atmospheric neutrino data}, Phys.Rev. D71 (2005)
  093010.
\newblock \href {http://arxiv.org/abs/hep-ph/0502223}
  {\path{arXiv:hep-ph/0502223}}.

\bibitem{Gao:2013xoa}
J.~Gao, M.~Guzzi, J.~Huston, H.-L. Lai, Z.~Li, et~al., {The CT10 NNLO Global
  Analysis of QCD. }\href {http://arxiv.org/abs/1302.6246}
  {\path{arXiv:1302.6246}}.

\bibitem{Gribov:1984tu}
L.~Gribov, E.~Levin, M.~Ryskin, {Semihard Processes in QCD}, Phys.Rept. 100
  (1983) 1--150.

\bibitem{Quigg:1986mb}
C.~Quigg, M.~Reno, T.~Walker, {Interactions of Ultrahigh-Energy Neutrinos},
  Phys.Rev.Lett. 57 (1986) 774.

\bibitem{Reno:1987zf}
M.~Reno, C.~Quigg, {On the Detection of Ultrahigh-Energy Neutrinos}, Phys.Rev.
  D37 (1988) 657.

\bibitem{Gandhi:1995tf}
R.~Gandhi, C.~Quigg, M.~H. Reno, I.~Sarcevic, {Ultrahigh-energy neutrino
  interactions}, Astropart.Phys. 5 (1996) 81--110.
\newblock \href {http://arxiv.org/abs/hep-ph/9512364}
  {\path{arXiv:hep-ph/9512364}}.

\bibitem{Lipari:1991ut}
P.~Lipari, T.~Stanev, {Propagation of multi - TeV muons}, Phys.Rev. D44 (1991)
  3543--3554.

\bibitem{Dziewonski:1981xy}
A.~Dziewonski, D.~Anderson, {Preliminary reference earth model}, Phys.Earth
  Planet.Interiors 25 (1981) 297--356.

\bibitem{Honda:2011nf}
M.~Honda, T.~Kajita, K.~Kasahara, S.~Midorikawa, {Improvement of low energy
  atmospheric neutrino flux calculation using the JAM nuclear interaction
  model}, Phys.Rev. D83 (2011) 123001.
\newblock \href {http://arxiv.org/abs/1102.2688} {\path{arXiv:1102.2688}}.

\bibitem{Volkova:1980sw}
L.~Volkova, {Energy Spectra and Angular Distributions of Atmospheric
  Neutrinos}, Sov.J.Nucl.Phys. 31 (1980) 784--790.

\bibitem{Gondolo:1995fq}
P.~Gondolo, G.~Ingelman, M.~Thunman, {Charm production and high-energy
  atmospheric muon and neutrino fluxes}, Astropart.Phys. 5 (1996) 309--332.
\newblock \href {http://arxiv.org/abs/hep-ph/9505417}
  {\path{arXiv:hep-ph/9505417}}.

\bibitem{Alexandreas:1992ek}
D.~Alexandreas, D.~Berley, S.~Biller, G.~Dion, J.~Goodman, et~al., {Point
  source search techniques in ultrahigh-energy gamma-ray astronomy},
  Nucl.Instrum.Meth. A328 (1993) 570--577.

\bibitem{Junk:1999kv}
T.~Junk, {Confidence level computation for combining searches with small
  statistics}, Nucl.Instrum.Meth. A434 (1999) 435--443.
\newblock \href {http://arxiv.org/abs/hep-ex/9902006}
  {\path{arXiv:hep-ex/9902006}}.

\bibitem{Read:2000ru}
A.~L. Read, {Modified frequentist analysis of search results (The CL(s)
  method)}, CERN-OPEN-2000-205.

\bibitem{ATLAS:2011tau}
{Procedure for the LHC Higgs boson search combination in summer 2011},
  ATL-PHYS-PUB-2011-011, ATL-COM-PHYS-2011-818, CMS-NOTE-2011-005.

\bibitem{Beringer:1900zz}
J.~Beringer, et~al., {Review of Particle Physics}, Phys.Rev. D86 (2012) 010001.

\bibitem{Aartsen:2013uuv}
M.~Aartsen, et~al., {Search for time-independent neutrino emission from
  astrophysical sources with 3 years of IceCube data}\href
  {http://arxiv.org/abs/1307.6669} {\path{arXiv:1307.6669}}.

\bibitem{Apel:2013dga}
W.~Apel, J.~Arteaga, L.~Bähren, K.~Bekk, M.~Bertaina, et~al., {Thunderstorm
  Observations by Air-Shower Radio Antenna Arrays. }\href
  {http://arxiv.org/abs/1303.7068} {\path{arXiv:1303.7068}}.

\bibitem{IceCube:2013wda}
M.~Aartsen, et~al., {Measurement of the cosmic ray energy spectrum with
  IceTop-73. }\href {http://arxiv.org/abs/1307.3795} {\path{arXiv:1307.3795}}.

\bibitem{Razzaque:2013uoa}
S.~Razzaque, {The Galactic Center Origin of a Subset of IceCube Neutrino
  Events}, Phys.Rev. D88 (2013) 081302.
\newblock \href {http://arxiv.org/abs/1309.2756} {\path{arXiv:1309.2756}}.

\bibitem{Aartsen:2012gka}
M.~Aartsen, et~al., {Search for Galactic PeV Gamma Rays with the IceCube
  Neutrino Observatory}, Phys.Rev. D87 (2013) 062002.
\newblock \href {http://arxiv.org/abs/1210.7992} {\path{arXiv:1210.7992}}.

\bibitem{Ahlers:2013xia}
M.~Ahlers, K.~Murase, {Probing the Galactic Origin of the IceCube Excess with
  Gamma-Rays. }\href {http://arxiv.org/abs/1309.4077} {\path{arXiv:1309.4077}}.

\end{thebibliography}


\end{document}